\documentclass[fleqn,usenatbib]{mnras}

\usepackage{newtxtext,newtxmath}

\usepackage[T1]{fontenc}

\DeclareRobustCommand{\VAN}[3]{#2}
\let\VANthebibliography\thebibliography
\def\thebibliography{\DeclareRobustCommand{\VAN}[3]{##3}\VANthebibliography}

\usepackage{graphicx}	
\usepackage{amsmath}	
\usepackage{gensymb}
\usepackage{booktabs}
\usepackage{xcolor}
\usepackage{float}

\title[New method for transient detection in radio images]{A new method for short duration transient detection in radio images: Searching for transient sources in MeerKAT data of NGC 5068}

\author[Fijma et al.]{
S. Fijma$^{1}$,\thanks{E-mail: s.c.fijma@uva.nl}
A. Rowlinson$^{1,2}$,
R.A.M.J. Wijers$^{1}$,
I. de Ruiter$^{1}$,
W.J.G. de Blok$^{2,3,4}$,
S. Chastain$^{5}$, 
\newauthor
A.J. van der Horst$^{5}$,
Z.S. Meyers$^{6, 7, 1}$,
K. van der Meulen$^{1}$,
R. Fender$^{8,3}$,
P.A. Woudt$^{3}$,
A. Andersson$^{8}$,
\newauthor
A. Zijlstra${^9}$,
J. Healy$^{2}$,
F. M. Maccagni$^{2,10}$
\\
$^{1}$Anton Pannekoek Institute for Astronomy, University of Amsterdam, Science Park 904, 1098 XH, Amsterdam, the Netherlands\\
$^{2}$ASTRON, the Netherlands Institute for Radio Astronomy, Oude Hoogeveensedijk 4, 7991 PD Dwingeloo, the Netherlands\\
$^{3}$Department of Astronomy, University of Cape Town, Private Bag X3, Rondebosch 7701, South Africa\\
$^{4}$Kapteyn Astronomical Institute, University of Groningen, PO Box 800, 9700 AV Groningen, The Netherlands\\
$^{5}$Department of Physics, The George Washington University, 725 21st Street NW, Washington, DC 20052, USA\\
$^{6}$Deutsches Elektronen-Synchrotron DESY, Platanenallee 6, 15738 Zeuthen, Germany\\
$^{7}$Erlangen Center for Astroparticle Physics (ECAP), Friedrich-Alexander-Universität Erlangen-Nuremberg, 91058 Erlangen, Germany\\
$^{8}$Astrophysics, Department of Physics, University of Oxford, Denys Wilkinson Building, Keble Road, Oxford OX1 3RH, UK\\
${^9}$Jodrell Bank Centre for Astrophysics, Department of Physics and Astronomy, University of Manchester, Manchester, M13 9PL, UK\\
$^{10}$INAF - Osservatorio Astronomico di Cagliari, Via della Scienza 5, I-09047 Selargius (CA), Italy
}

\date{Accepted XXX. Received YYY; in original form ZZZ}

\pubyear{2023}

\begin{document}
\label{firstpage}
\pagerange{\pageref{firstpage}--\pageref{lastpage}}
\maketitle

\begin{abstract}
Transient surveys are a vital tool in exploring the dynamic universe, with radio transients acting as beacons for explosive and highly energetic astrophysical phenomena. However, performing commensal transient surveys using radio imaging can require a significant amount of computing power, data storage and time. With the instrumentation available to us, and with new and exciting radio interferometers in development, it is essential that we develop efficient methods to probe the radio transient sky. 
In this paper, we present results from an commensal short duration transient survey, on time-scales of 8 seconds, 128 seconds and 1 hour, using data from the MeerKAT radio telescope. The dataset used was obtained as part of a galaxy observing campaign, and we focus on the field of NGC 5068. We present a quick, wide-field imaging strategy to enable fast imaging of large datasets, and develop methods to efficiently filter detected transient candidates. 
No transient candidates were identified on the time-scales of 8 seconds, 128 seconds and 1 hour, leading to competitive limits on the transient surface densities of $6.7\,{\times}\,10^{-5}$ deg$^{-1}$, $1.1\,{\times}\,10^{-3}$ deg$^{-1}$, and $3.2\,{\times}\,10^{-2}$ deg$^{-1}$  at sensitivities of 56.4 mJy, 19.2 mJy, and 3.9 mJy for the respective time-scales. We find one possible candidate that could be associated with a stellar flare, that was rejected due to strict image quality control. 
Further short time-scale radio observations of this candidate could give definite results to its origin.
\end{abstract}




\section{Introduction}
\label{sec:introduction}

Exploring the radio transient sky has proven invaluable in studying highly-energetic and/or explosive astrophysical phenomena. 
Searching for transient radio emission allows us to constrain the population of known source types, discover new types \citep[such as Fast Radio Bursts (FRBs); see][]{lorimer2007}, investigate the associated sources for these events and study the resulting kinetic feedback in the local environment \citep[e.g.][]{2016mks..confE..13F..Fender2016..thunderkat}.

With the continuing development of more sensitive radio telescopes with excellent uv- and sky-coverage, for example MeerKAT \citep{camilo2018}, the Australian SKA Precursor \citep[ASKAP;][]{johnston2008} and the future Square Kilometer Array \citep[SKA;][]{braun2015}, 
the search for transient sources in the radio sky has become of great interest to many researchers. So far, many types of transient source have been discovered, varying on time-scales from milliseconds to years. These can generally be divided in two categories: incoherent and coherent radio transients \cite[see e.g.][for an overview]{2015MNRAS.446.3687P..Pietka2015}. Incoherent radio transients are proposed to emit through synchrotron radiation, and occur on time scales above one second. Coherent radio transients are proposed to emit radiation in phase through Microwave Amplification by Stimulated Emission of Radiation (MASER) or coherent synchrotron emission, and typically occur on short time scales of order seconds or less. We can study these sources using radio observations in the image-domain, where the time scale we can study is limited by the integration time (typically a few seconds), or using time series analysis, where a time resolution of microseconds or less is achievable.
Furthermore, we can identify both transient and variable sources, where we define transient sources in this work as sources that are newly detected during the observations.

Many teams have searched for radio transient sources in large surveys, using both imaging and time series analysis. 
In time series analysis, the main focus is on searching and describing radio pulsars, as well as on the detection of FRBs to determine their properties and likely progenitor systems \citep[for a recent review see ][]{petroff2022}. In the image plane, a number of teams have searched large data sets for transient sources on time-scales from minutes to years 
\citep[e.g.][see also Table \ref{table:transientSDs}]{driessen2020, driessen2022, 2022MNRAS.516.5972W..Wang2022,2023MNRAS.tmp.1665W..Wang2023, 2022MNRAS.510.3794D..Dobie2022, 2023MNRAS.519.4684D..Dobie2023, 2023MNRAS.tmp.1391..Andersson2023}.
These studies are referred to as unbiased or commensal transient searches as they do not target known transient sources. At 1.4 GHz, commensal transient surveys have determined that radio transients are typically rare, but several have been identified \citep[such as gamma-ray burst (GRB) afterglows, stars, pulsars, galaxies and active-galactic nuclei (AGN), see e.g.][]{levinson2002,thyagarajan2011,aoki2014,murphy2021,andersson2022}. 

In this work, we focus on short duration transient sources that are found via image-plane surveys. 
Interesting transient sources have been detected on sub-minute time-scales by a few surveys. For example, \cite{hurleywalker2022} identified a periodic radio transient with pulses of duration 30--60 seconds in snapshot images produced from survey data obtained by the Murchison Widefield Array at around $150$ MHz. 
Additionally, the detection of a 76 second period pulsar in time series analysis of MeerKAT observations and its subsequent detection in simultaneous 8 second snapshot images from MeerKAT \citep[as part of observations by the MeerTRAP\footnote{\url{https://www.meertrap.org/}}  and ThunderKAT\footnote{\url{https://http://www.thunderkat.uct.ac.za/}} projects;][]{caleb2022}, shows the high potential of using short duration snapshot images to find new types of transient source such as slowly spinning neutron stars.

To search for rare transient sources, one would ideally survey the entire sky at all times so as to not miss any events. For example, the AARTFAAC All Sky Monitor achieves this for a large part of the low-frequency radio sky \citep[10-90 MHz;][]{2016JAI.....541008P..Prasad2016..AARTFAAC,kuiack2021} by generating radio images of 1 s in real time. Additionally, at high time resolution, the Canadian Hydrogen Intensity Mapping Experiment \citep[CHIME;][]{2018ApJ...863...48C..CHIME2018} and the Survey for Transient Astronomical Radio Emission \citep[STARE2;][]{2020PASP..132c4202B..Bochenek2020.STARE2} survey a large part of the sky in real time for very short time scale transients (${\approx}1$ ms) at frequencies of around 400-800 MHz and 1280-1530 MHz, respectively. However, achieving this with image-plane surveys at 1.4 GHz is extremely challenging, due to the smaller fields of view of the facilities and the prohibitive computational costs of calibrating and imaging the data.
Therefore, commensal transient surveys typically utilise archival data or data obtained by key survey projects obtained by the facilities \citep[e.g.][]{bell2011,hancock2016}.

As radio transients are likely to be rare, large sky areas need to be processed to maximise the chance of detection. To cover as much sky area as possible, one would need observations covering a large field of view, or large numbers of fields with smaller sky area. Moreover, sensitive data is preferred, to be able to detect transient sources within a large range of peak luminosities. However, creating robust images on short time slices and performing a transient survey on this data still requires a significant amount of computing power, data storage and time. 

Imaging software like WSClean \citep{2014MNRAS.444..606O..Offringa2014..WSClean} enables efficient wide field imaging, and allows us to create time slices or snapshots in single observations. Using these time slices, transients can be studied on time scales spanning the length of the observation (typically a few hours) to the integration time of the observation (typically a few seconds). However, despite the efficiencies in WSClean, several of the steps are highly time consuming. Cleaning the sources in the images is particularly slow as it is an iterative process. Additionally, software is used to detect and monitor sources, for example the {\sc LOFAR Transients Pipeline} \citep[{\sc TraP};][]{2015A&C....11...25S..Swinbank2015}. The time such processing pipelines take to analyse the images is highly dependent upon the number of sources per image. Thus, further optimisation is required to significantly reduce the time it takes to conduct short duration snapshot transient surveys.

In this paper, we present results from a MeerKAT short duration transient survey, on time-scales of 8 seconds up to 1 hour, conducted as part of the ThunderKAT large survey project to detect and monitor transient sources \citep{2016mks..confE..13F..Fender2016..thunderkat}. We present a quick, wide-field imaging strategy to enable imaging of large datasets. As we focus on the search for short duration radio transients, we aim not to monitor known sources in the field as this reduces the time required for subsequent analysis pipelines. In Section \ref{sec:data}, we outline the data used in this paper and the calibration strategy. In Section \ref{sec:imaging}, we outline the fast imaging strategy we used and its motivation. Section \ref{sec:analysis} outlines the image analysis steps including the detection and filtering of transient candidates. Section \ref{sec:surface_density} compares our results to previous surveys conducted at 1.4 GHz.

\vspace{-0.2cm}
\section{Data}
\label{sec:data}

The data used in this work were obtained as part of the large survey project MeerKAT HI Observations of Nearby Galactic Objects - Observing Southern Emitters \citep[MHONGOOSE;][]{2016mks..confE...7D..deBlok2016}, which aims to perform a deep survey of the neutral hydrogen distribution in a sample of galaxies using the MeerKAT radio telescope. We are able to perform a transient survey on their data as part of the ThunderKAT commensal search programme \citep{2016mks..confE..13F..Fender2016..thunderkat}. We study an observation of NGC 5068, which was made as part of the early commissioning of MHONGOOSE on April 28th, 2020. 
In this work we only use one observation to extensively test our developed method and to perform a detailed analysis on the obtained transient candidates.  

NGC 5068 was observed with a total bandwidth of 3.27 MHz, a central frequency of 1.417 GHz, and 1001 frequency channels. The total length of the observation is $2.67\,{\times}\,10^{4}$ s or ${\approx}7.41$ hours, and the visibilities were recorded every 8 seconds. J1939-6342 was used as the primary flux and bandpass calibrator, and J1311-2216 was used as the secondary phase calibrator. The target and secondary calibrator were observed sequentially, with 12 minutes spent on source followed by 2 minutes spent on the secondary calibrator. The resulting total time on target is ${\approx}1.96\,{\times}\,10^{4}$ s or ${\approx}5.45$ hours.

The data have been reduced by the MHONGOOSE team as described by \cite{2019A&A...628A.122S..Serra2019} and \cite{2020A&A...643A.147D..deBlok2020}, using a pre-release version of the {\sc CARACal} pipeline \citep{2020ASPC..527..635J..Jozsa2020, 2019A&A...628A.122S..Serra2019}. 
The pipeline uses components from various astronomy packages as one single execution. It first flags calibrator and target data using {\sc AOFlagger} \citep[][]{offringa2010}. 
Then it performs a cross-calibration step using the primary and secondary calibrator, and a self-calibration step. We use this calibrated dataset to perform our fast imaging strategy.

\section{Fast imaging strategy}
\label{sec:imaging}

As mentioned in Section \ref{sec:introduction}, we aim to perform a computationally efficient transient survey. In this section, we outline the methods used to achieve this and an analysis of the quality of the images obtained.

\subsection{Continuum subtraction}
A continuum subtraction step is applied, where the self-calibrated continuum model is subtracted from the data in the uv-plane \citep{2020A&A...643A.147D..deBlok2020}. 
Any sources that are not present in the deep image are not detected throughout the full observation and are therefore not subtracted from the visibilities. Thus, the only sources remaining in the snapshot images are transient sources.

This continuum extraction step speeds up analysis twofold. Firstly, the removal of sources from the visibilities leads to faster processing by WSClean. Secondly, by only keeping sources that are transient candidates the subsequent image analysis steps using {\sc TraP} are significantly faster.

An additional positive consequence of this step is that the final snapshot image rms noise is significantly lower than for the non-subtracted images and is approaching thermal noise. This is because the confusion noise component of the rms in the snapshot image is being drastically reduced. Therefore, we are also maximising the sensitivity of the transient survey we are conducting.

\subsection{Snapshot imaging}
\label{sec:subtraction}

As briefly motivated in Section \ref{sec:introduction}, we image the data using WSClean \citep{2014MNRAS.444..606O..Offringa2014..WSClean}. To study transient sources at the time scales of 1 hour, 128 seconds and 8 seconds, we create 6, 153 and 2454 time slices of the data, respectively. We create these time slices using the \texttt{IntervalsOut} command. The images per time slice are created at 3 frequency bands (1.416, 1.417, 1.418 GHz), together with a combined multi-frequency summed (MFS) image. 

Using the \texttt{FitBeam} command when imaging the full data set, we fit the shape of the point-spread function (PSF) and use this to determine the size and position angle of the restoring beam, which corresponds to $7.79"\,{\times}\,6.68"$ for the restoring beam size in the continuum subtracted data.
Based on the restoring beam size, we set the \texttt{PixelScale} setting to $2"$, so that the beam would cover around 9 to 10 pixels in the image. The image size was based on the amount of data that could be efficiently imaged, i.e. where the run time was not excessive, which is $5120\,{\times}\,5120$ pixels or $2.87\degree\,{\times}\,2.87\degree$ in size. The weighting scheme is set to Briggs weighting, with the robustness parameter set to 0. 

As mentioned in Section \ref{sec:introduction}, the standard imaging step of cleaning is very slow due to it being an iterative process. As we are imaging continuum subtracted data, the only sources should be transients and we expect the majority of images to contain no sources. Therefore, we decide to skip the time-consuming cleaning process and use the '{dirty}' images in our further analysis. Additionally, we do not perform a primary beam correction as this also slows down processing.

\subsection{Image quality control}
\label{sec:image_quality_control}
Once we have created time sliced images for the 1 hour, 128 second and 8 second time scales for our data, we evaluate the quality of the images by measuring the average rms noise variation in the inner one-eight of the images. This is measured for each image when using the {\sc TraP}, which will be discussed in detail in the next Section. 
It should be noted that there is some faint emission in the centre of the image as a result of an incomplete subtraction of the continuum emission of galaxy NGC 5068. However, as this only covers the inner one-twenty-fifth of the image and only faint emission remains, we can use the average rms noise calculated in the inner one-eight of the images.

\begin{figure}
    \centering
    \includegraphics[width=\columnwidth]{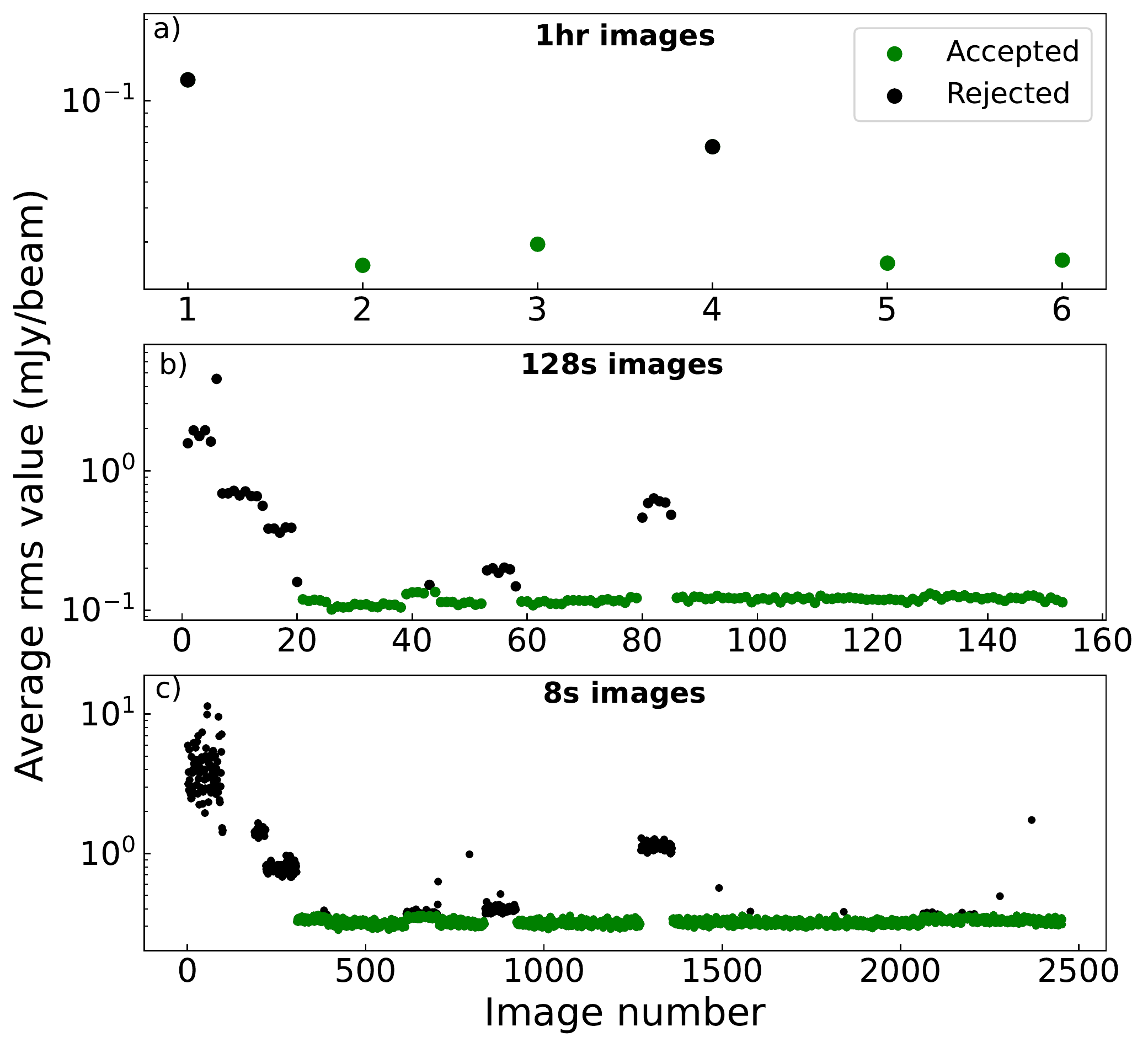}
    \caption{The average rms values for all images for each time scale. The top panel (a) shows the average rms values for the 1 hour time scale images, the middle panel (b) for the 128 second time scale images, and the bottom panel (c) for the 8 second time scale images. The average rms value variation is measured in the inner one-eight of each image by {\sc TraP}. The green data points indicate accepted images for the transient survey, and the black data points indicate rejected images.} 
    \label{fig:RMS}
\end{figure}

In Figure \ref{fig:RMS} we show the average rms noise for the MFS images per time scale. Based on this distribution, we iteratively reject images where the rms noise deviates more than $3\sigma$ from the mean. These images likely have high values for the average rms noise due to poor uv-coverage, a turbulent ionosphere, radio-frequency interference (RFI) and/or poor calibration of the data, as discussed by e.g. \cite{2016MNRAS.458.3506R..Rowlinson2016}.
33 out of 153 images (or $21.6\%$) in the 128 second time scale are rejected, and 528 out of 2454 images (or $21.5\%$) in the 8 second time scale are rejected. 
Furthermore, we manually reject 2 out of 6 images (or $33.3\%$) in the 1 hour time scale, as they are evidently poor quality images but are not removed using the $3\sigma$ clip due to the low number of images.
The remaining images are used for our transient survey.

\section{Analysis}
\label{sec:analysis}

\subsection{Detecting transient candidates}
\label{sec:detecting}

To detect transient sources in the images, we use the {\sc TraP} \citep[Release 4.0;][]{2015A&C....11...25S..Swinbank2015}.
{\sc TraP} has several settings we can tweak to optimise our transient searches, but most of the default {\sc TraP} settings are appropriate for our survey. 
We set \texttt{force\_beam} to \texttt{True}, so {\sc TraP} sets all source fits assuming that all sources have the size, shape and orientation of the restoring beam and keeps these parameters constant. 
The \texttt{new\_source\_sigma\_margin} parameter determines the margin of error (as a multiple of the rms of the previous best image) for which a new candidate is considered a false positive resulting from fluctuations around the detection threshold. We set this to 0, so we can consider all transient candidates resulting from our survey. The \texttt{extraction\_radius} is the radius in pixels in which {\sc TraP} will search for candidates. We evaluate the noise distribution throughout the image, and the edges were not considerably noisier than the centre. 
The radius is usually set to around the primary beam size in commensal transient surveys, as the telescope is most sensitive here. For the MeerKAT telescope, the HWHM of the telescope beam at 1.42 GHz is ${\approx}0.52\degree$ \citep{2020ApJ...888...61M..Mauch2020}.
Because of the noise distribution in our images, and to test out our method for a larger area of the images, we decided to include a large extraction radius in our survey, and to set this at 2500 pixels (or around $1.4\degree$). This can be reconsidered for future surveys to improve efficiency.

Lastly, we determine the \texttt{detection\_threshold} setting, which is expressed as a multiple of the rms noise. To limit the amount of false positive detections, we calculate the \texttt{detection\_threshold} such that we would expect < 1 false positive detections caused by noise. In Section \ref{sec:data} we mentioned that we chose the pixel scale such that the restoring beam size would be ${\approx}10$ pixels. A false positive detection could then be the result of noise when around 10 pixels have values in Jy/beam exceeding the detection threshold. We can expect this assuming that the noise in a radio image follows a Gaussian distribution and that the noise is random, caused by e.g. contributions of sky and receiver thermal noise, so the noise pixel values are independent of each other. 

\begin{figure}
    \centering
    \includegraphics[width=\columnwidth]{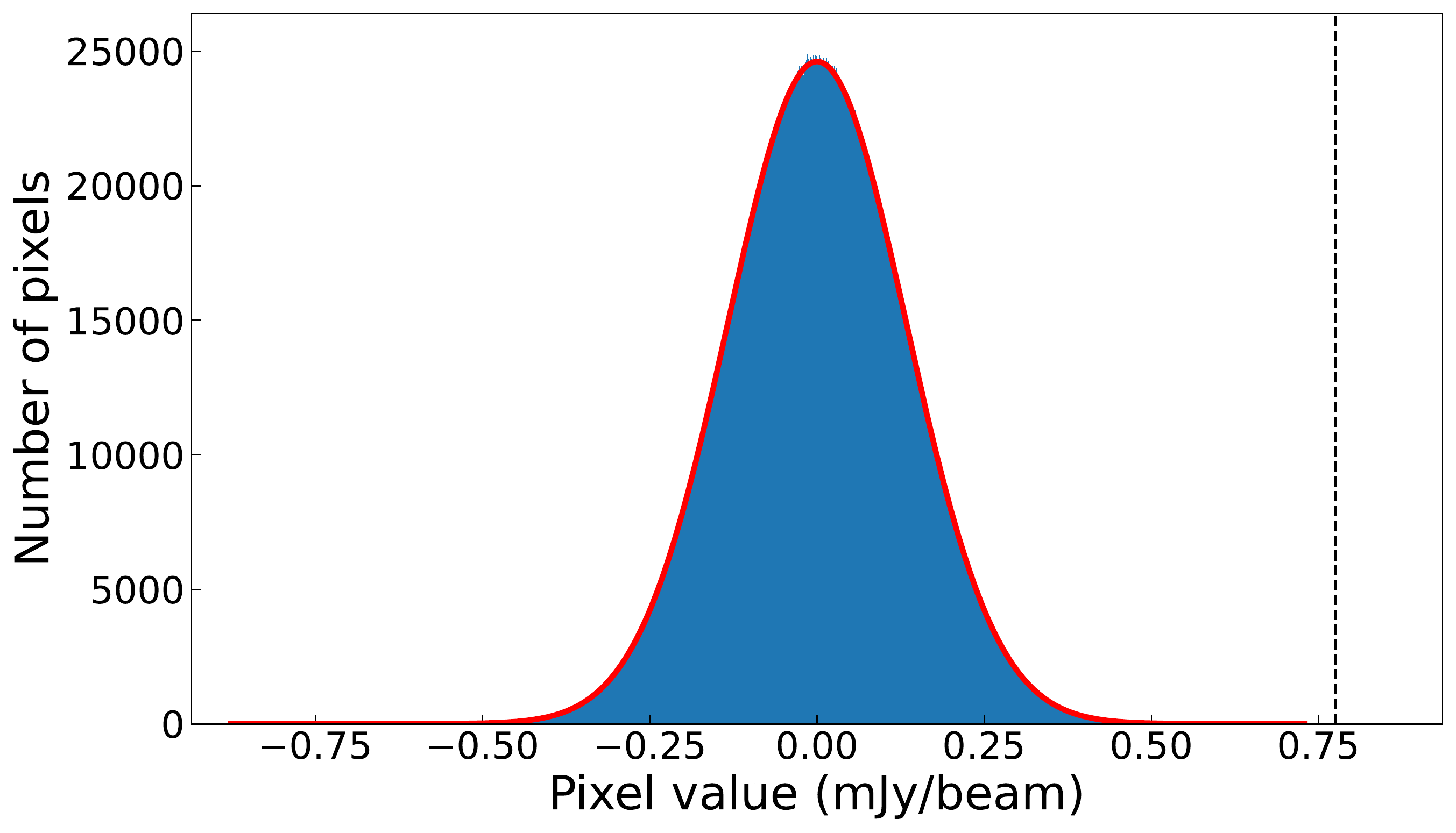}
    \caption{The pixel value distribution of an example image (image 39) for the 128 second time scale shown in blue, with a fitted Gaussian distribution shown in red. The corresponding detection threshold for this data set, at $5.8\sigma$, is indicated with a black dashed line.
    As the image is continuum-subtracted, the mean of the pixel value distribution is close to zero.} 
    \label{fig:gaussian}
\end{figure}

A Gaussian pixel distribution may not always be applicable for radio images. Noise can be correlated in between different images and within individual images \citep[e.g.][]{2016MNRAS.458.3506R..Rowlinson2016}, which changes the distribution. Moreover, calibration artefacts and side lobes present in the image also affect the distribution. Therefore, we fit a Gaussian distribution to a random sample of the 1 hour, 128 second and 8 second images, to see if this assumption would hold for our data. We find that a Gaussian distribution fits our data well. we show an example in Figure \ref{fig:gaussian}. 

We calculate the detection threshold based on the probability that we encounter 10 pixels exceeding the detection threshold for the total number of pixels per data set for the 1 hour, 128 second and 8 second time scales. This can be expressed as $Pr(X \leq x) = 1 - \frac{1}{N}$. N is the total number of beams per data set, or the total number of pixels in the data set divided by the number of pixels in the restoring beam. We calculate the value of $x$ expressed in $\sigma$ (i.e. the threshold) for this probability using the the percent point function (or inverse cumulative distribution function) for the Gaussian distributions of the images per time scale using the \texttt{scipy stats norm ppf} functionality \citep{2020NatMe..17..261V..Virtanen2020.scipy}. 
It is not likely for 10 randomly distributed noise pixels exceeding the detection threshold to all appear in the same location in the image, and in the shape and size of the restoring beam. Therefore, this method is mainly used to get a good approximation for the detection threshold. 
In Table \ref{tab:threshold} we show the calculated detection thresholds per time scale data set, and the total number of candidates we obtain for each of the data sets. The thresholds were calculated for a random sample of accepted images per time scale, which gave consistent values for the threshold to 3 significant digits for each time scale. The thresholds were then scaled to all pixels for each time scale data set.

\begin{table}
\resizebox{\columnwidth}{!}{%
\begin{tabular}{@{}lcccc@{}}
\toprule
Time scale  & Total number & Number of images & Detection & Number of  \\ 
            & of images    & after QC & threshold & candidates \\ \midrule
1 hour      & 6       &     4    & $5.3 \sigma$          & 17                   \\
128 seconds & 153     &     120    & $5.8 \sigma$          & 8                    \\
8 seconds   & 2454    &     1926    & $6.3 \sigma$          & 4                    \\ \bottomrule
\end{tabular}%
}
\caption{The number of images used for the transient survey for each time scale. The total number of images are filtered using quality control (QC) steps described in Section \ref{sec:image_quality_control}. Using the detection threshold calculated based on expecting $<\,1$ false positive per time scale (see Section \ref{sec:detecting}), we present the initial number of transient candidates detected in the filtered images.}
\label{tab:threshold}
\end{table}

\subsection{Filtering transient candidates}
\label{sec:filtering}

Besides false positive detections caused by noise, more types of false positive are expected in radio transient surveys. \cite{2022MNRAS.509.5018G..Gourdji2021} describes some categories of detected transient candidates, including: 
\begin{itemize}
    \item Imaging artefacts from sidelobes around bright sources
    \item Extended sources with fit parameters that differ significantly between images preventing them from being associated with one another by {\sc TraP}
    \item Faint sources detected around the detection threshold in an image, but that were undetected in at least the first image due to higher local rms
\end{itemize}
To separate false positive detections from true astrophysical detections, we employ a number of filtering steps to our lists of transient candidates per time scale data set. 

As described in Section \ref{sec:detecting}, we initially filter false positive detections caused by noise using the calculated \texttt{detection\_threshold} per time scale data set. In Table \ref{tab:threshold} we show the total amount of candidates detected using the detection thresholds. 

The source finder used in TraP \citep[PySE][]{2018A&C....23...92C..Carbone2018..PySE} calculates the noise in an image by dividing the source extraction region into a grid. 
\cite{2016MNRAS.458.3506R..Rowlinson2016} find that the local rms is not accurately calculated close to the extraction radius by {\sc TraP}, because {\sc TraP} does not model the rms noise beyond the source extraction region. Therefore, we reject any candidates with a distance greater than 2460 pixels from the centre of the image. 
Because TraP uses bilinear interpolation in calculating the background and rms maps \citep{2015A&C....11...25S..Swinbank2015}, this results in inaccurate calculations and thus can result in false positive detections close to the extraction radius. The background size used to calculate the local rms is set to $50\,{\times}\,50$ pixels. We expect the rms to be accurately calculated for sources in the middle of this box when the entire box is included in the extraction radius. Therefore, we filter any sources closer than half the diagonal of the background box to the extraction radius, which we round up to 40 pixels. As the extraction radius for all transient searches is set to 2500 pixels, we choose to reject any candidates with a distance greater than 2460 pixels from the centre of the image. This step resulted in 4 out of 4 candidates in the 8 second time scale data to be rejected. No candidates in other time scales were rejected using this step. 

We also perform a visual inspection of the transient candidates. The aforementioned types of false positive detection are easily distinguished by eye from potential astrophysical transient sources \cite[see e.g. Figure A1 in][]{2022MNRAS.509.5018G..Gourdji2021}. Furthermore, we also identify an additional false positive detection type caused by detected remnants from the applied continuum subtraction. These appear like bright points in noisy areas in the location of bright sources in the original data. 
For our visual inspection, we only accept transient candidates in the shape of the restoring beam and do not fit the description of the false positive detection types. 
Moreover, we filter any candidates that are present in the image of the full observation, as the remaining sources would be artefacts of the continuum subtraction, and as we are only interested in detecting transient sources. 
8 out of 8 candidates in the 128 second time scale data are rejected using this step, and 12 out of 17 candidates the 1 hour time scale data.

Once we obtain promising candidates from the visual inspection, we proceed to use the {\sc TraP} monitoring functionality to monitor the location of the transient candidates in all images of the 1 hour, 128 second and 8 second time scales, to obtain light curves of the candidates on these time scales. 
This allows us to check if variations in the light curves are significant, or a result of the signal-to-noise ratio fluctuating around the detection threshold. We show an example of a rejected candidate in Figure \ref{fig:lc_detec_threshold}. It also allows us to characterise the behaviour of the candidate sources and the area in which they are located. This results in 3 out of 5 candidates in the 1 hour time scale data to be rejected. 

If the variations are significant compared to the noise measurements before and after the transient behaviour, we proceed to monitor noise surrounding the candidates. Some detections can result from a local increase in the noise, ionospheric effects, poor calibration or high RFI. Therefore, we create and check the light curves of 5 random points surrounding the candidates. We monitor these points in the image where the candidate was initially detected, as well as 10 images before and 10 images after this image, to analyse the noise close in time to the detection. Furthermore, we generate the random points at least a distance of 5 times the size of the restoring beam from the candidates, to limit the effects of imaging artefacts and residual dirty beam effects close to the source, see e.g. \cite{2021MNRAS.508.2412D..deRuiter2021}. We also select the points within a distance of $2’$ (approximately 15 times the major axis of the restoring beam), so that the points are representative for monitoring noise specifically close to the candidate. If the statistical errors for the integrated flux density for one or more of the random (noise) points overlaps with the statistical errors for the integrated flux density of the candidate during the observed variation in the light curve, i.e. if the peak of the detection does not deviate significantly from the noise, we reject the candidate. We show an example in Figure \ref{fig:lc_noise_peak}. 
We also reject any candidates where the noise shows a similar evolution in flux as the flux of the candidate. 
This results in 2 out of 2 candidates in the 1 hour time scale data to be rejected. 

In summary, no transient candidates pass all of the filtering criteria for this survey.

\subsection{Rejected transient candidate}
\label{sec:rejectedcandidate}

One candidate passed the filtering strategies in a pilot survey of the data, when the quality checks for the images were less conservative. 
However, after performing the image rms analysis for our final survey, the image in which we detected the candidate (image 57, see Figure \ref{fig:RMS}) was rejected. 
Although this candidate is not considered as a potential astrophysical transient source for our survey, we do believe it would be interesting to consider if the data are reprocessed or in future observations of the field. 
Moreover, we were also able to test our methodology of how a transient candidate from this type of dataset would be processed.

\begin{figure*}
    \centering
    \includegraphics[width=\textwidth]{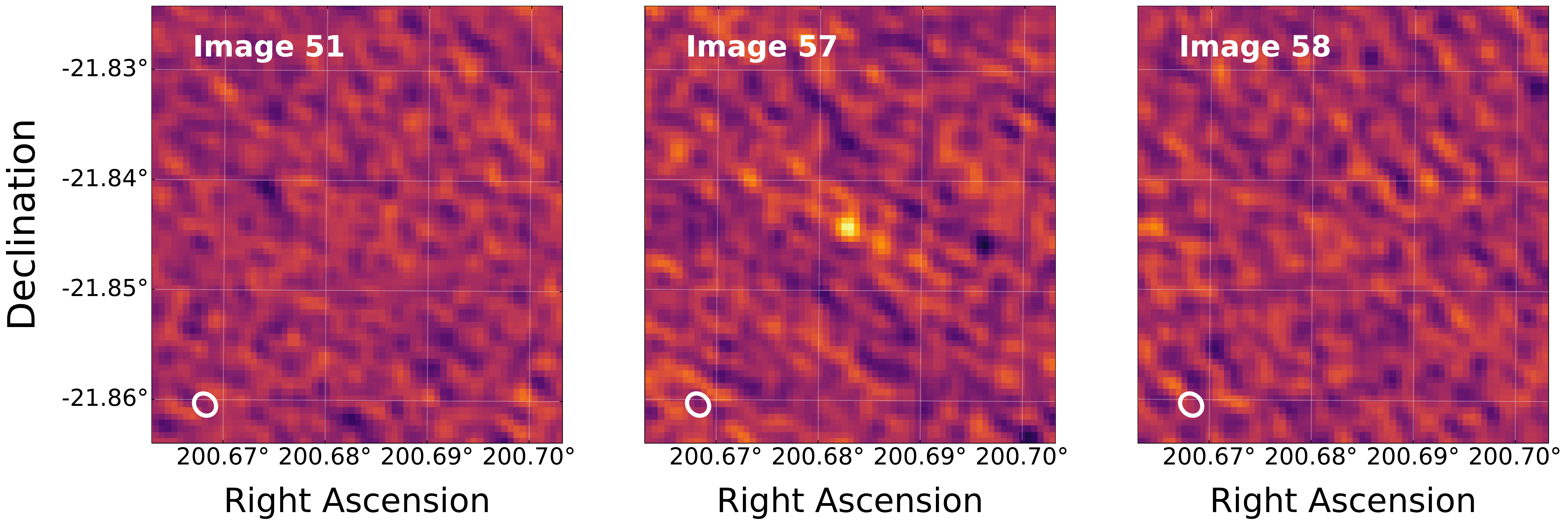}
    \caption{
    MeerKAT images for the 128 second time scale showing the location of the rejected transient candidate, each with a dimension of $2.4'\,{\times}\,2.4'$. The middle panel shows the image in which the candidate was detected, image 57, with a signal-to-noise ratio of $6.02\sigma$. The left panel shows image 51, which is the first accepted image (based on quality control, see Section \ref{sec:image_quality_control}) preceeding image 57. The right panel shows image image 58, which is the first accepted image following image 57.
    The contour of the shape and size of the restoring beam as determined for full data set by {\sc TraP} is shown in white in the lower left corner.}
    \label{fig:image_candidate}
\end{figure*}

\begin{figure}
    \centering
    \includegraphics[width=\columnwidth]{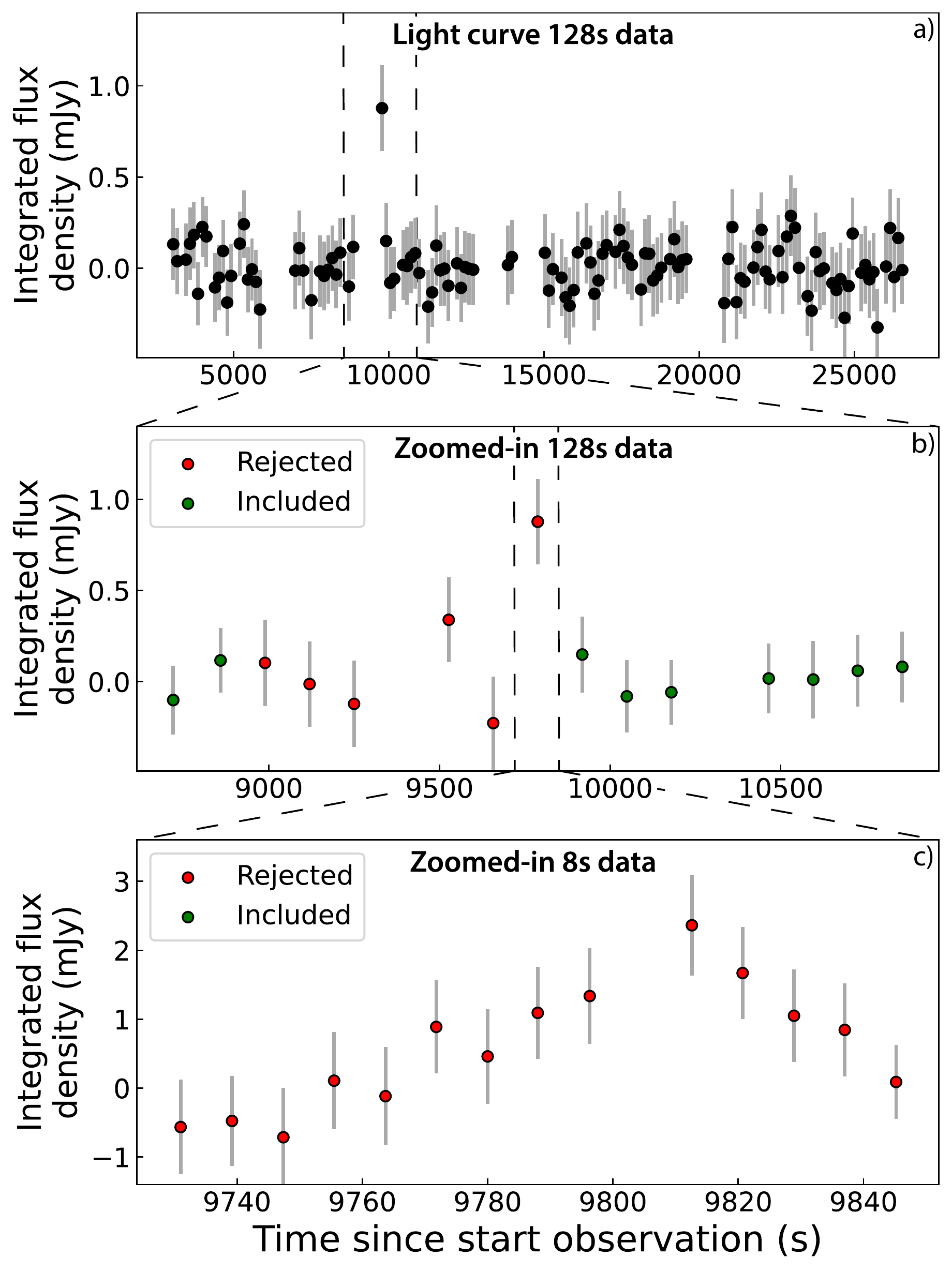}
    \caption{The integrated flux density in mJy measured by {\sc TraP} in each image at the location of the rejected transient candidate, plotted against the time since the start of the observation. Panel (a) The light curve for all images for the full 128 second time scale data set, showing a clear peak in 1 image, where the candidate was detected. Panel (b) The light curve of 15 images around the detection of the candidate of the 128 second time scale data set. The images rejected by our quality control are shown in red, and the accepted images are shown in green. Panel (c) The light curve of the 8 second time scale images around the time of the detection in the 128 second time scale image.}
    \label{fig:rejectedcandidate}
\end{figure}

The candidate was detected in the 128 second time scale data set at right ascension (R.A.) $200.683\degree$ and declination (DEC) $-21.844\degree$
with a signal-to-noise ratio of $6.02\sigma$ and an integrated flux density of $0.90\,{\pm}\,0.23$ mJy without primary beam correction. 
In Figure \ref{fig:image_candidate} we show the transient candidate before, during and after its detection in the 128 second time scale data set. In Figure \ref{fig:rejectedcandidate} we show the light curves of the transient candidate, using the 128 second data where the candidate was initially detected, and using the 8 second images to improve the time resolution of the light curve. We find a clear peak in the 8 second light curve at the time of the detection, which lasts around ${\approx}64$ seconds. Thus, the source is detected in multiple snapshots and in images with different time slicing. This adds strength to the candidate as it is then unlikely to be caused by features in the radio images such as side lobes or incomplete source subtractions. This is because imaging artefacts tend to appear in different locations within the images when using different uv-coverage (i.e. changes in integration time or observing frequency).

The signal-to-noise ratio of the transient candidate exceeds the detection threshold for the 128 second data of $5.8\sigma$. When studying the image, the candidate appears point-like and has the shape of the restoring beam (as determined by {\sc TraP} for the full data set), with no excessive noise or artefacts surrounding the candidate. 
The light curve shows a clear variation from the flux in other images, and the candidate is within 2460 pixels of the centre of the image. Furthermore, we monitored random points surrounding the candidate to study the noise, and we find that the light curves of the random noise points do not follow the same trend as the light curve of the candidate. 

Lastly, we also check the images of the different frequency bands. RFI could also result in a false positive detection in radio images, but RFI generally spans a more limited frequency range than the frequency range of our observation \citep[see e.g.][]{offringa2010}. 
Therefore, if this is a false positive detection resulting from RFI, we expect to only identify emission in one or two of the frequency band images. We identify point-like emission at the location of the candidate in all three frequency band images.
Moreover, RFI would be visible across the image, as a satellite or plane would leave multiple candidates in a line across the image, \citep[e.g.][]{kuiack2021}. 
In this case, we would also expect more features in the noise, which we do not identify for this candidate. Therefore, we deem the scenario of RFI causing a false positive detection for this candidate to be unlikely. 

Our method for the transient survey involves analysing the dirty continuum-subtracted data. To study the candidate further, we created cleaned images of the original non-continuum subtracted data, to see if we detect the transient candidate data in these images as well. 
We were not able to recover this transient candidate in the cleaned non-continuum subtracted data. However, we find that the (re)imaged cleaned time slice images from the non-continuum subtracted data have higher rms noise than the original time sliced dirty images using continuum subtracted data. This is based on analysing the pixel value distribution (we show an example in Figure \ref{fig:noise_cleaned_images}), as well as studying the integrated flux density of random points in the images. 
As noted in Section \ref{sec:subtraction}, without the continuum subtraction we are not able to probe as deep as in our original survey due to confusion noise. 

We perform a catalogue search to see if we can find an associated source for the candidate. One source is detected by PanSTARRS \citep{2016arXiv161205560C..Chambers2016..PanSTARRS}, Gaia DR3 \citep{2016A&A...595A...1G..Gaia2016} 
and 2MASS \citep{2006AJ....131.1163S..Skrutskie2006..2MASS}, at R.A. $200.68070925(6) \degree$ and DEC $-21.84227554(4)\degree$ for Gaia DR3. This is within $10"$ of the position of the transient candidate, which is the default systematic positional uncertainty measured with TraP. The PanSTARRS observed magnitudes are
$g\,{=}\,19.735(11)$, $r\,{=}\,18.801(6)$ and $i\,{=}\,18.362(4)$. These give colours of $r{-}i\,{=}\,0.439(7)$ and $g{-}r\,{=}\,0.934(12)$, 
which are consistent with a K-type star on the main sequence \citep[see Figure 1 in][]{finlator2000}. Therefore, the duration of the transient and its optical association are consistent with this transient being a stellar flare. 

Calculations in \cite{Yu2021} of expected flare star densities predict one flare from an M-dwarf brighter than 1 mJy (at 5 GHz) over 400 square degrees in an 8 hour integration, giving a predicted density of $2.5 \times 10^{-3}$ per square degree for an 8 hour observation. This gives a probability of around 2\% to detect an M-dwarf flare, with the fraction of flares decreasing for earlier types, i.e. for the proposed K-type star. Moreover, based on photometric data of the associated optical source from the Catalina Real-Time Transient Survey Data Release 2 \citep[CRTS DR2;][]{Drake2009}, no flaring behaviour is found. Follow-up research using short time-scale radio and optical observations are thus needed to further study this transient and its potential optical counterpart.

Because of the filtering method we employ for the rejection of low quality images, we cannot justify including this candidate as a potential astrophysical transient source for our survey. 
The average rms for the image where this candidate was initially detected, image 57, is $0.15$ mJy/beam, so $4.5\sigma$ removed from the mean of $0.12$ mJy/beam. The previous 5 images (52 to 56) were also rejected using this method; however, the average rms values for these images deviated around $10$ to $13\sigma$ from the mean. The decision to include or exclude image 57 is therefore somewhat of an edge case. 
Generally, we find that applying an image filtering method is necessary for our survey, as including low quality images results in detecting a large number of false positive candidates. Nonetheless, from this result it appears that our image filtering method could be too conservative for our transient survey, meaning that we might miss real astrophysical transient sources in the data. 

We tried different methods for filtering images based on the rms. We attempted to calculate the value for the mean and standard deviation of the entire image data set, and only reject images deviating more than $3\sigma$ from the mean once, instead of using iterative clipping. We also tried iteratively clipping images deviating $4\sigma$ from the mean instead of $3\sigma$. However, we found that our current method was more appropriate, because otherwise we would include evidently low quality images as well.  

This transient candidate, although rejected in the final analysis, suggests that there may indeed be exciting transients to be found on these time-scales and - given that this was only a small dataset - they may even be plentiful. If this transient is indeed a stellar flare, then we expect to find more flares from this source in the full MHONGOOSE observations of this field.

\section{Transient Surface Density}
\label{sec:surface_density}

\begin{figure*}
    \centering
    \includegraphics[width=0.9\textwidth]{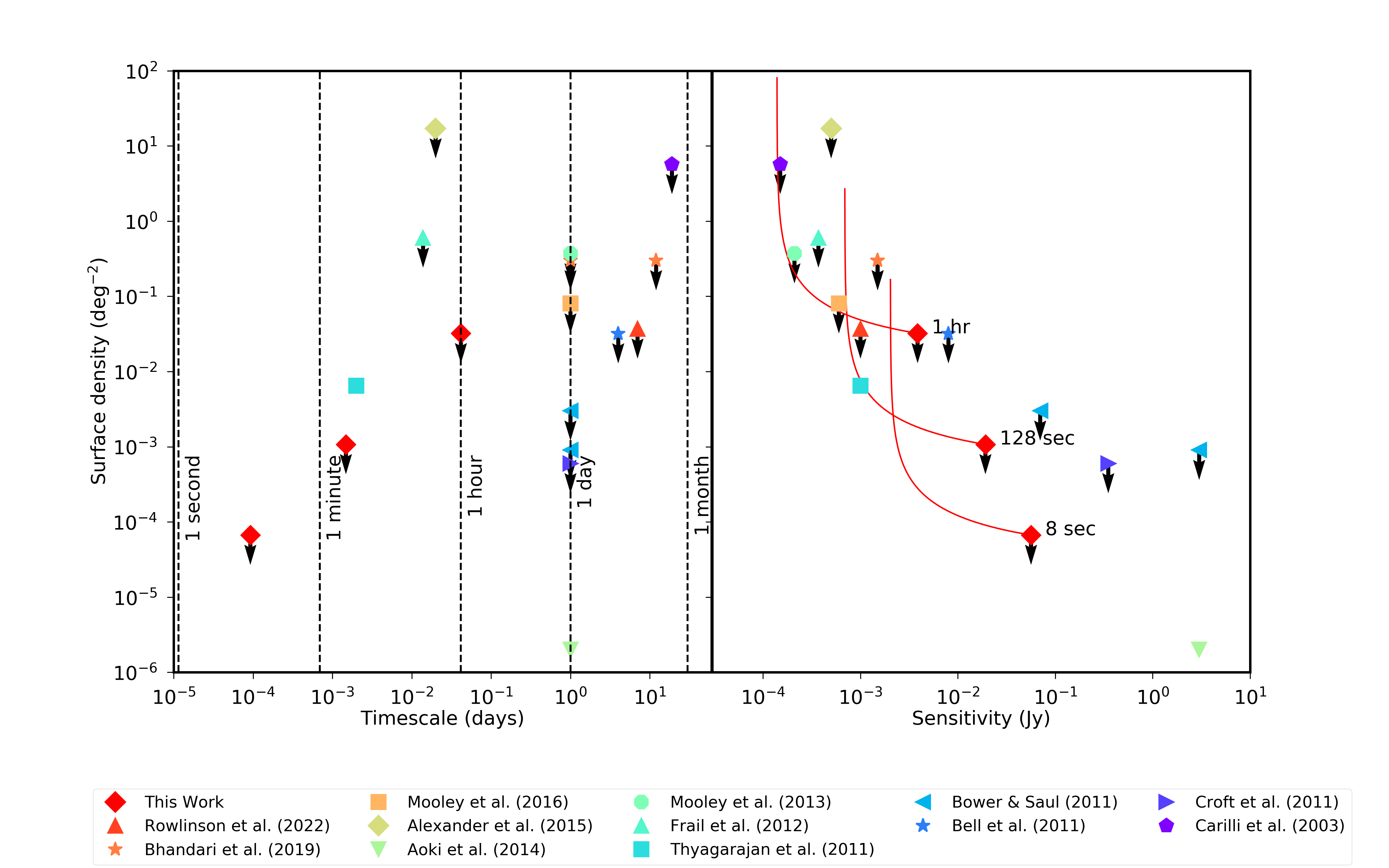}
    \caption{This figure shows the transient surface density versus the time-scale of the transient (on the left) and the faintest detectable transient (the sensitivity, on the right). The key shows the surveys included in this plot; these surveys are at 1.4 GHz and are for transient time-scales of less than 1 month. The red diamonds are the three surveys presented in this paper, on time-scales of 8 seconds, 128 seconds and 1 hour. The red curved lines show the sensitivity as a function of the surface density for each survey, taking into account the changing sensitivity across the images caused by the primary beam.}
    \label{fig:surfaceDensityPlot}
\end{figure*}

In this section, we compare our transient survey to previous transient surveys at ${\approx}1.4$ GHz \citep[given in Table \ref{table:transientSDs} and obtained from][]{mooley2016}\footnote{\url{http://www.tauceti.caltech.edu/kunal/radio-transient-surveys/index.html}}. These surveys are typically compared using the transient surface density, which is the number of transients detected per square degree surveyed, and the sensitivity of the survey, which is defined as the faintest detectable transient in the survey.

For each image, we extracted sources out to a radius of 1.4 degrees giving a sky area of ${\approx}5.9$ square degrees per image. 
Following the method developed in \cite{2016MNRAS.458.3506R..Rowlinson2016}, we can calculate the typical sensitivity as a function of radius by dividing the sensitivity in the centre of the image by the primary beam response. The MeerKAT primary beam main lobe is well modelled using a cosine aperture taper \citep{condon2016} and this model is provided using {\sc katbeam}\footnote{\url{https://github.com/ska-sa/katbeam}}. For each observing frequency, we use {\sc katbeam} to model the 2D primary beam response and take a cross-section through the beam response from a radius of 0 to the search radius in degrees giving ${\rm beam}(r)$. We then obtain the sensitivity as a function of radius using:
\begin{equation}
    {\rm Sensitivity}(r) = \frac{{\rm Detection~threshold}\,{\cdot}\,{\rm RMS}_c}{{\rm beam}(r)}, \label{eqn:sensitivity}
\end{equation}
where the detection threshold in $\sigma$ is given in Table \ref{tab:threshold} and the ${\rm RMS}_c$ is the average rms noise in the centre of the images for a given search time-scale. We calculate this sensitivity curve for each image and then combine the results by summing the total area searched across all images for given sensitivity bins.

To calculate the total transient surface density probed, as a function of sensitivity, we first calculate the area enclosed within each radius and multiply it by the number of images surveyed at each time-scale. For a $95\%$ confidence limit, following e.g. \cite{bell2014,2016MNRAS.458.3506R..Rowlinson2016}, the transient surface density is given by:
\begin{equation}
    \rho(r) = \frac{3}{A(r)\,{\cdot}\,N_{\rm imgs}}, \label{eqn:surfDens}
\end{equation}
where $A(r)$ is the area probed in one image out to a radius of $r$ and $N_{\rm imgs}$ is the number of images probed for each time-scale surveyed. Combining this equation with the areas calculated using equation \ref{eqn:sensitivity} for given sensitivity bins, we can calculate the transient surface density as a function of the transient search sensitivity.

Using equations \ref{eqn:sensitivity} and \ref{eqn:surfDens}, we plot the transient surface density limits as a function of the sensitivity in Figure \ref{fig:surfaceDensityPlot} for each of the three time-scales probed in this survey (red curves). The red diamonds show the lowest surface density probed (i.e. the largest sky area surveyed) at the worst sensitivity and these values are given in Table \ref{table:transientSDs}.

On 8 second time-scales, we find an upper limit on the transient surface density of $6.7\,{\times}\,10^{-5}$ deg$^{-1}$ at a sensitivity of 56.4 mJy. 
This is the one of the first commensal transient surveys of its kind on 8 second time-scales at 1.4 GHz, together with other MeerKAT studies (Chastain et al. 2023, in prep).
On these time-scales, we might expect to find sources such as slow pulsars. Recently a slow pulsar, J0901-4046, was detected using high time domain and imaging data obtained as part of the MeerTRAP and ThunderKAT projects \citep{caleb2022}. J0901-4046 has a 76 second spin period and pulses from this unusual pulsar were clearly detected in 8 second snapshot images created using the MeerKAT imaging data and localised to arc second precision. Additionally, a periodic radio transient has been discovered in snapshot imaging at around 150 MHz by the Murchison Widefield Array; with a spin period of 18.18 minutes and pulses of 30-60 second duration \citep{hurleywalker2022}. \cite{hurleywalker2022} suggest that this source may be a radio magnetar due to its unusual properties. The pulsar and candidate magnetar detections demonstrate that these sources would be detectable in commensal transient searches using MeerKAT observations, such as the dataset presented in this paper. However, this field is at high Galactic latitude (b = 41 degrees) and, hence, we expect the number density of pulsars to be very low in this field.

On 128 second time-scales, we find an upper limit on the transient surface density of $1.1\,{\times}\,10^{-3}$ deg$^{-1}$ at a sensitivity of 19.2 mJy. As shown in Figure \ref{fig:surfaceDensityPlot} and outlined above, this limit is best represented by a curved line taking into account the improved sensitivity in the inner regions of the images. This curved line passes through a data point obtained by \cite{thyagarajan2011}, who conducted an commensal transient survey on a similar time-scale of 3 minutes. The observations processed by \cite{thyagarajan2011} comprised of 65,000 images from the FIRST survey \citep{becker1995} at 1.4 GHz. They detected 57 sources identified as being transient, with most having no archival association. As the line representing our survey passes through the data point of \cite{thyagarajan2011}, we can deduce that we would expect to detect ${\approx}1$ transient source in our observations. Although it has several caveats, the source presented in Section \ref{sec:rejectedcandidate} would be consistent with our expectations from \cite{thyagarajan2011}. We also note, that with just 7.41 hours of MeerKAT data, we are able to match the constraints obtained using the entire FIRST survey. Two other transient sources have been detected on the several minutes time-scales at lower radio frequencies. The first source is GCRT 1745, known as the `Galactic Burper', and emits flares of duration ${\approx}10$ minutes with a periodicity of 77 minutes \citep{hyman2005}. The nature of the `Galactic Burper' remains unknown, with a leading theory being a magnetar. As the Galactic Burper is located near the Galactic Centre, it is unlikely that the observations presented in this paper would detect a similar object due to their high Galactic latitude. The second source at low radio frequencies was found well off the Galactic Plane in imaging observations obtained by LOFAR \citep{stewart2016}. This source had a duration of a few minutes, a poorly constrained flux density of 15-25 Jy and was detected at 60 MHz. The nature of this source also remains unknown. Due to the significant difference in observing frequency and unknown source spectrum, it is unclear if transients of this type would be detectable in our observations at 1.4 GHz. 
The sources found by \citep{hyman2005} and \citep{stewart2016} were detected on lower frequencies than our survey, which is why they are not plotted in Figure \ref{eqn:surfDens}.

On 1 hour time-scales, we find an upper limit on the transient surface density of $3.2\,{\times}\,10^{-2}$ deg$^{-1}$ at a sensitivity of 3.9 mJy. From the sensitivity -- surface density red curve presented in Figure \ref{fig:surfaceDensityPlot}, we note that this survey constitutes the deepest transient survey on 1 hour time-scales to date. The nearest comparable surveys \citep[with durations of ${\lesssim}1$ hour;][]{alexander2015,frail2012} have surface densities 1--2 orders of magnitude higher than the survey presented in this paper at a comparable sensitivity. On this time-scale and the 128 seconds time-scale, we might expect to detect flare stars in our local region of our Galaxy \citep[see e.g.][]{driessen2022,andersson2022}. Detectable Galactic flare stars would be expected to be somewhat isotropically distributed across the sky due to their proximity to Earth and, hence, the high Galactic latitude of the field presented in this paper should not significantly impact upon the rates of detection.

\section{Discussion}
\label{sec:discussion}

In this work, we present a blind radio transient survey on around 7 hours of MeerKAT data. We have used a search radius of 1.4 degrees in each image, and considered three time scales to study; namely 8 seconds, 128 seconds and, 1 hour. A total sky area of around $1.13\,{\times}\,10^4$, $7.04\,{\times}\,10^2$, 23.5 square degrees has been surveyed at a sensitivity of 56.4, 19.2, and 3.9 mJy for the 8 s, 128 s, and 1 hr time scales, respectively. 

We have developed a new imaging and filtering strategy to efficiently perform the transient survey, and we find that this method is faster and deeper than searching standard snapshots. The method is effective for finding faint short-duration transient phenomena in radio images, which makes it promising for future surveys. 
However, at the moment it cannot be used to monitor variability. As interesting variable sources are currently being found through new MeerKAT commensal transient surveys \citep[e.g.][]{rowlinson2022, driessen2022}, 
this would be interesting to explore in the future. 

Once transient candidates are found, the data can be re-imaged to obtain properties such as the accurate flux density. It is important to note that the candidates should be sufficiently bright to achieve this, as the re-imaged data can probe higher values for the rms noise than the original images, as discussed in Section \ref{sec:rejectedcandidate}. Consequently, we were unable to recover the rejected transient candidate presented in this section. 
However, alternative methods can be used to determine if detected candidates are true astrophysical transients. For example, by probing other radio observations of the same field, as well as through looking for counterparts in other observing bands. As the rejected transient candidate could potentially be a flare star, one could probe other MHONGOOSE observations of the same field, to confirm if transient emission is detected at the location of the candidate.

The one transient candidate detected in our transient survey was rejected because of our strict quality control settings. We tried a few methods to adjust our quality control settings, as discussed in Section \ref{sec:rejectedcandidate}, but this resulted in including evidently low quality images as well. As the quality control method was developed for standard snap shot images, this might need adapting for similar data and methods in future work. For example, by evaluating the local image quality for transient candidates as well, instead of only evaluating the quality for the inner one-eight of the image. 

Moreover, we are able to place competitive limits on transient rates for the time-scales of 1 hour ($3.2\,{\times}\,10^{-2}$ deg$^{-1}$ at a sensitivity of 3.9 mJy) and 128 seconds ($1.1\,{\times}\,10^{-3}$ deg$^{-1}$ at a sensitivity of 19.2 mJy), through analysing only 7 hours of data. 
When comparing our findings to surveys at similar time scales, we would expect to find one transient at the 128 second time scale. 
If the rejected transient candidate presented in Section \ref{sec:rejectedcandidate} is a true astrophysical transient, our results would be consistent with findings in other surveys. Finally, we also conducted one of the first commensal transient surveys of its kind on 8 second time-scales at 1.4 GHz.

By combining radio data from new facilities with excellent uv-coverage, and new rapid imaging and search strategies such as the methods presented in this work, we will be able to probe radio transients on time scales from seconds to hours to high sensitivity. 
Facilities such as MeerKAT, ASKAP, and the highly anticipated SKA, will allow us to make significant progress in studying short duration image-plane radio transients.

\section*{Acknowledgements}

We thank M. Sipior for assisting with the virtual machine lofproc, and M. Heemskerk for help with the Helios cluster. 
We thank M. Kuiack, K. Gourdji, J. Meijn, J. van den Eijnden, N. Degenaar for useful discussions. 
AR acknowledges funding from the NWO Aspasia grant (number: 015.016.033).
IdR acknowledges support through the project CORTEX (NWA.1160.18.316) of the research programme NWA-ORC which is (partly) financed by the Dutch Research Council (NWO).
PAW acknowledges support from the National Research Foundation (grant number 129359) and the University of Cape Town.
This project has received funding from the European Research Council (ERC) under the European Unions Horizon 2020 research and innovation program grant agreement No. 882793, project name MeerGas.)
The MeerKAT telescope is operated by the South African Radio Astronomy Observatory (SARAO), which is a facility of the National Research Foundation, an agency of the Department of Science and Innovation. We would like to thank the operators, SARAO staff and ThunderKAT Large Survey Project team. 

The Pan-STARRS1 Surveys (PS1) and the PS1 public science archive have been made possible through contributions by the Institute for Astronomy, the University of Hawaii, the Pan-STARRS Project Office, the Max-Planck Society and its participating institutes, the Max Planck Institute for Astronomy, Heidelberg and the Max Planck Institute for Extraterrestrial Physics, Garching, The Johns Hopkins University, Durham University, the University of Edinburgh, the Queen's University Belfast, the Harvard-Smithsonian Center for Astrophysics, the Las Cumbres Observatory Global Telescope Network Incorporated, the National Central University of Taiwan, the Space Telescope Science Institute, the National Aeronautics and Space Administration under Grant No. NNX08AR22G issued through the Planetary Science Division of the NASA Science Mission Directorate, the National Science Foundation Grant No. AST-1238877, the University of Maryland, Eotvos Lorand University (ELTE), the Los Alamos National Laboratory, and the Gordon and Betty Moore Foundation.
This work has made use of data from the European Space Agency (ESA) mission {\it Gaia} (\url{https://www.cosmos.esa.int/gaia}), processed by the {\it Gaia} Data Processing and Analysis Consortium (DPAC, \url{https://www.cosmos.esa.int/web/gaia/dpac/consortium}). Funding for the DPAC has been provided by national institutions, in particular the institutions participating in the {\it Gaia} Multilateral Agreement.
This publication makes use of data products from the Two Micron All Sky Survey, which is a joint project of the University of Massachusetts and the Infrared Processing and Analysis Center/California Institute of Technology, funded by the National Aeronautics and Space Administration and the National Science Foundation.
The CSS survey is funded by the National Aeronautics and Space Administration under Grant No. NNG05GF22G issued through the Science Mission Directorate Near-Earth Objects Observations Program. The CRTS survey is supported by the U.S.~National Science Foundation under grants AST-0909182 and AST-1313422.

\section*{Data Availability}

The data and scripts underlying this article will be available in Zenodo at DOI: 10.5281/zenodo.7883761 upon publication. 
The MeerKAT data used are available in the MeerKAT data archive at \url{https://archivesarao.ac.za/}. The PanSTARRS fits images were obtained from \url{https://panstarrs.stsci.edu}.




\begin{thebibliography}{}
\makeatletter
\relax
\def\mn@urlcharsother{\let\do\@makeother \do\$\do\&\do\#\do\^\do\_\do\%\do\~}
\def\mn@doi{\begingroup\mn@urlcharsother \@ifnextchar [ {\mn@doi@}
  {\mn@doi@[]}}
\def\mn@doi@[#1]#2{\def\@tempa{#1}\ifx\@tempa\@empty \href
  {http://dx.doi.org/#2} {doi:#2}\else \href {http://dx.doi.org/#2} {#1}\fi
  \endgroup}
\def\mn@eprint#1#2{\mn@eprint@#1:#2::\@nil}
\def\mn@eprint@arXiv#1{\href {http://arxiv.org/abs/#1} {{\tt arXiv:#1}}}
\def\mn@eprint@dblp#1{\href {http://dblp.uni-trier.de/rec/bibtex/#1.xml}
  {dblp:#1}}
\def\mn@eprint@#1:#2:#3:#4\@nil{\def\@tempa {#1}\def\@tempb {#2}\def\@tempc
  {#3}\ifx \@tempc \@empty \let \@tempc \@tempb \let \@tempb \@tempa \fi \ifx
  \@tempb \@empty \def\@tempb {arXiv}\fi \@ifundefined
  {mn@eprint@\@tempb}{\@tempb:\@tempc}{\expandafter \expandafter \csname
  mn@eprint@\@tempb\endcsname \expandafter{\@tempc}}}

\bibitem[\protect\citeauthoryear{{Alexander}, {Soderberg}  \&
  {Chomiuk}}{{Alexander} et~al.}{2015}]{alexander2015}
{Alexander} K.~D.,  {Soderberg} A.~M.,   {Chomiuk} L.~B.,  2015, \mn@doi [\apj]
  {10.1088/0004-637X/806/1/106}, \href
  {https://ui.adsabs.harvard.edu/abs/2015ApJ...806..106A} {806, 106}

\bibitem[\protect\citeauthoryear{{Andersson} et~al.,}{{Andersson}
  et~al.}{2022}]{andersson2022}
{Andersson} A.,  et~al., 2022, \mn@doi [\mnras] {10.1093/mnras/stac1002}, \href
  {https://ui.adsabs.harvard.edu/abs/2022MNRAS.513.3482A} {513, 3482}

\bibitem[\protect\citeauthoryear{{Andersson} et~al.,}{{Andersson}
  et~al.}{2023}]{2023MNRAS.tmp.1391..Andersson2023}
{Andersson} A.,  et~al., 2023, \mn@doi [\mnras] {10.1093/mnras/stad1298}, \href
  {https://ui.adsabs.harvard.edu/abs/2023MNRAS.tmp.1391A} {}

\bibitem[\protect\citeauthoryear{{Aoki} et~al.,}{{Aoki}
  et~al.}{2014}]{aoki2014}
{Aoki} T.,  et~al., 2014, \mn@doi [\apj] {10.1088/0004-637X/781/1/10}, \href
  {https://ui.adsabs.harvard.edu/abs/2014ApJ...781...10A} {781, 10}

\bibitem[\protect\citeauthoryear{{Becker}, {White}  \& {Helfand}}{{Becker}
  et~al.}{1995}]{becker1995}
{Becker} R.~H.,  {White} R.~L.,   {Helfand} D.~J.,  1995, \mn@doi [\apj]
  {10.1086/176166}, \href
  {https://ui.adsabs.harvard.edu/abs/1995ApJ...450..559B} {450, 559}

\bibitem[\protect\citeauthoryear{{Bell} et~al.,}{{Bell}
  et~al.}{2011}]{bell2011}
{Bell} M.~E.,  et~al., 2011, \mn@doi [\mnras]
  {10.1111/j.1365-2966.2011.18631.x}, \href
  {https://ui.adsabs.harvard.edu/abs/2011MNRAS.415....2B} {415, 2}

\bibitem[\protect\citeauthoryear{{Bell} et~al.,}{{Bell}
  et~al.}{2014}]{bell2014}
{Bell} M.~E.,  et~al., 2014, \mn@doi [\mnras] {10.1093/mnras/stt2200}, \href
  {https://ui.adsabs.harvard.edu/abs/2014MNRAS.438..352B} {438, 352}

\bibitem[\protect\citeauthoryear{{Bhandari} et~al.,}{{Bhandari}
  et~al.}{2018}]{bhandari2018}
{Bhandari} S.,  et~al., 2018, \mn@doi [\mnras] {10.1093/mnras/sty1157}, \href
  {https://ui.adsabs.harvard.edu/abs/2018MNRAS.478.1784B} {478, 1784}

\bibitem[\protect\citeauthoryear{{Bochenek}, {McKenna}, {Belov}, {Kocz},
  {Kulkarni}, {Lamb}, {Ravi}  \& {Woody}}{{Bochenek}
  et~al.}{2020}]{2020PASP..132c4202B..Bochenek2020.STARE2}
{Bochenek} C.~D.,  {McKenna} D.~L.,  {Belov} K.~V.,  {Kocz} J.,  {Kulkarni}
  S.~R.,  {Lamb} J.,  {Ravi} V.,   {Woody} D.,  2020, \mn@doi [\pasp]
  {10.1088/1538-3873/ab63b3}, \href
  {https://ui.adsabs.harvard.edu/abs/2020PASP..132c4202B} {132, 034202}

\bibitem[\protect\citeauthoryear{{Bower} \& {Saul}}{{Bower} \&
  {Saul}}{2011}]{bower2011}
{Bower} G.~C.,  {Saul} D.,  2011, \mn@doi [\apjl]
  {10.1088/2041-8205/728/1/L14}, \href
  {https://ui.adsabs.harvard.edu/abs/2011ApJ...728L..14B} {728, L14}

\bibitem[\protect\citeauthoryear{{Braun}, {Bourke}, {Green}, {Keane}  \&
  {Wagg}}{{Braun} et~al.}{2015}]{braun2015}
{Braun} R.,  {Bourke} T.,  {Green} J.~A.,  {Keane} E.,   {Wagg} J.,  2015, in
  Advancing Astrophysics with the Square Kilometre Array (AASKA14). p.~174

\bibitem[\protect\citeauthoryear{{CHIME/FRB Collaboration} et~al.,}{{CHIME/FRB
  Collaboration} et~al.}{2018}]{2018ApJ...863...48C..CHIME2018}
{CHIME/FRB Collaboration} et~al., 2018, \mn@doi [\apj]
  {10.3847/1538-4357/aad188}, \href
  {https://ui.adsabs.harvard.edu/abs/2018ApJ...863...48C} {863, 48}

\bibitem[\protect\citeauthoryear{{Caleb} et~al.,}{{Caleb}
  et~al.}{2022}]{caleb2022}
{Caleb} M.,  et~al., 2022, \mn@doi [Nature Astronomy]
  {10.1038/s41550-022-01688-x}, \href
  {https://ui.adsabs.harvard.edu/abs/2022NatAs...6..828C} {6, 828}

\bibitem[\protect\citeauthoryear{{Camilo}}{{Camilo}}{2018}]{camilo2018}
{Camilo} F.,  2018, \mn@doi [Nature Astronomy] {10.1038/s41550-018-0516-y},
  \href {https://ui.adsabs.harvard.edu/abs/2018NatAs...2..594C} {2, 594}

\bibitem[\protect\citeauthoryear{{Carbone} et~al.,}{{Carbone}
  et~al.}{2018}]{2018A&C....23...92C..Carbone2018..PySE}
{Carbone} D.,  et~al., 2018, \mn@doi [Astronomy and Computing]
  {10.1016/j.ascom.2018.02.003}, \href
  {https://ui.adsabs.harvard.edu/abs/2018A&C....23...92C} {23, 92}

\bibitem[\protect\citeauthoryear{{Carilli}, {Ivison}  \& {Frail}}{{Carilli}
  et~al.}{2003}]{carilli2003}
{Carilli} C.~L.,  {Ivison} R.~J.,   {Frail} D.~A.,  2003, \mn@doi [\apj]
  {10.1086/375005}, \href
  {https://ui.adsabs.harvard.edu/abs/2003ApJ...590..192C} {590, 192}

\bibitem[\protect\citeauthoryear{{Chambers} et~al.,}{{Chambers}
  et~al.}{2016}]{2016arXiv161205560C..Chambers2016..PanSTARRS}
{Chambers} K.~C.,  et~al., 2016, arXiv e-prints, \href
  {https://ui.adsabs.harvard.edu/abs/2016arXiv161205560C} {p. arXiv:1612.05560}

\bibitem[\protect\citeauthoryear{{Condon} \& {Ransom}}{{Condon} \&
  {Ransom}}{2016}]{condon2016}
{Condon} J.~J.,  {Ransom} S.~M.,  2016, {Essential Radio Astronomy}

\bibitem[\protect\citeauthoryear{{Croft}, {Bower}, {Keating}, {Law}, {Whysong},
  {Williams}  \& {Wright}}{{Croft} et~al.}{2011}]{croft2011}
{Croft} S.,  {Bower} G.~C.,  {Keating} G.,  {Law} C.,  {Whysong} D.,
  {Williams} P. K.~G.,   {Wright} M.,  2011, \mn@doi [\apj]
  {10.1088/0004-637X/731/1/34}, \href
  {https://ui.adsabs.harvard.edu/abs/2011ApJ...731...34C} {731, 34}

\bibitem[\protect\citeauthoryear{{Dobie} et~al.,}{{Dobie}
  et~al.}{2022}]{2022MNRAS.510.3794D..Dobie2022}
{Dobie} D.,  et~al., 2022, \mn@doi [\mnras] {10.1093/mnras/stab3628}, \href
  {https://ui.adsabs.harvard.edu/abs/2022MNRAS.510.3794D} {510, 3794}

\bibitem[\protect\citeauthoryear{{Dobie} et~al.,}{{Dobie}
  et~al.}{2023}]{2023MNRAS.519.4684D..Dobie2023}
{Dobie} D.,  et~al., 2023, \mn@doi [\mnras] {10.1093/mnras/stac3731}, \href
  {https://ui.adsabs.harvard.edu/abs/2023MNRAS.519.4684D} {519, 4684}

\bibitem[\protect\citeauthoryear{{Drake} et~al.,}{{Drake}
  et~al.}{2009}]{Drake2009}
{Drake} A.~J.,  et~al., 2009, \mn@doi [\apj] {10.1088/0004-637X/696/1/870},
  \href {https://ui.adsabs.harvard.edu/abs/2009ApJ...696..870D} {696, 870}

\bibitem[\protect\citeauthoryear{{Driessen} et~al.,}{{Driessen}
  et~al.}{2020}]{driessen2020}
{Driessen} L.~N.,  et~al., 2020, \mn@doi [\mnras] {10.1093/mnras/stz3027},
  \href {https://ui.adsabs.harvard.edu/abs/2020MNRAS.491..560D} {491, 560}

\bibitem[\protect\citeauthoryear{{Driessen} et~al.,}{{Driessen}
  et~al.}{2022}]{driessen2022}
{Driessen} L.~N.,  et~al., 2022, \mn@doi [\mnras] {10.1093/mnras/stac756},
  \href {https://ui.adsabs.harvard.edu/abs/2022MNRAS.512.5037D} {512, 5037}

\bibitem[\protect\citeauthoryear{{Fender} et~al.,}{{Fender}
  et~al.}{2016}]{2016mks..confE..13F..Fender2016..thunderkat}
{Fender} R.,  et~al., 2016, in MeerKAT Science: On the Pathway to the SKA.
  p.~13 (\mn@eprint {arXiv} {1711.04132})

\bibitem[\protect\citeauthoryear{{Finlator} et~al.,}{{Finlator}
  et~al.}{2000}]{finlator2000}
{Finlator} K.,  et~al., 2000, \mn@doi [\aj] {10.1086/316824}, \href
  {https://ui.adsabs.harvard.edu/abs/2000AJ....120.2615F} {120, 2615}

\bibitem[\protect\citeauthoryear{{Frail}, {Kulkarni}, {Ofek}, {Bower}  \&
  {Nakar}}{{Frail} et~al.}{2012}]{frail2012}
{Frail} D.~A.,  {Kulkarni} S.~R.,  {Ofek} E.~O.,  {Bower} G.~C.,   {Nakar} E.,
  2012, \mn@doi [\apj] {10.1088/0004-637X/747/1/70}, \href
  {https://ui.adsabs.harvard.edu/abs/2012ApJ...747...70F} {747, 70}

\bibitem[\protect\citeauthoryear{{Gaia Collaboration} et~al.,}{{Gaia
  Collaboration} et~al.}{2016}]{2016A&A...595A...1G..Gaia2016}
{Gaia Collaboration} et~al., 2016, \mn@doi [\aap]
  {10.1051/0004-6361/201629272}, \href
  {https://ui.adsabs.harvard.edu/abs/2016A&A...595A...1G} {595, A1}

\bibitem[\protect\citeauthoryear{{Gourdji}, {Rowlinson}, {Wijers}, {Broderick},
  {Shulevski}  \& {Jonker}}{{Gourdji}
  et~al.}{2022}]{2022MNRAS.509.5018G..Gourdji2021}
{Gourdji} K.,  {Rowlinson} A.,  {Wijers} R.~A.~M.~J.,  {Broderick} J.~W.,
  {Shulevski} A.,   {Jonker} P.~G.,  2022, \mn@doi [\mnras]
  {10.1093/mnras/stab3197}, \href
  {https://ui.adsabs.harvard.edu/abs/2022MNRAS.509.5018G} {509, 5018}

\bibitem[\protect\citeauthoryear{{Hancock}, {Drury}, {Bell}, {Murphy}  \&
  {Gaensler}}{{Hancock} et~al.}{2016}]{hancock2016}
{Hancock} P.~J.,  {Drury} J.~A.,  {Bell} M.~E.,  {Murphy} T.,   {Gaensler}
  B.~M.,  2016, \mn@doi [\mnras] {10.1093/mnras/stw1486}, \href
  {https://ui.adsabs.harvard.edu/abs/2016MNRAS.461.3314H} {461, 3314}

\bibitem[\protect\citeauthoryear{{Hurley-Walker} et~al.,}{{Hurley-Walker}
  et~al.}{2022}]{hurleywalker2022}
{Hurley-Walker} N.,  et~al., 2022, \mn@doi [\nat] {10.1038/s41586-021-04272-x},
  \href {https://ui.adsabs.harvard.edu/abs/2022Natur.601..526H} {601, 526}

\bibitem[\protect\citeauthoryear{{Hyman}, {Lazio}, {Kassim}, {Ray}, {Markwardt}
   \& {Yusef-Zadeh}}{{Hyman} et~al.}{2005}]{hyman2005}
{Hyman} S.~D.,  {Lazio} T. J.~W.,  {Kassim} N.~E.,  {Ray} P.~S.,  {Markwardt}
  C.~B.,   {Yusef-Zadeh} F.,  2005, \mn@doi [\nat] {10.1038/nature03400}, \href
  {https://ui.adsabs.harvard.edu/abs/2005Natur.434...50H} {434, 50}

\bibitem[\protect\citeauthoryear{{Johnston} et~al.,}{{Johnston}
  et~al.}{2008}]{johnston2008}
{Johnston} S.,  et~al., 2008, \mn@doi [Experimental Astronomy]
  {10.1007/s10686-008-9124-7}, \href
  {https://ui.adsabs.harvard.edu/abs/2008ExA....22..151J} {22, 151}

\bibitem[\protect\citeauthoryear{{J{\'o}zsa} et~al.,}{{J{\'o}zsa}
  et~al.}{2020}]{2020ASPC..527..635J..Jozsa2020}
{J{\'o}zsa} G.~I.~G.,  et~al., 2020, in {Pizzo} R.,  {Deul} E.~R.,  {Mol}
  J.~D.,  {de Plaa} J.,   {Verkouter} H.,  eds,  Astronomical Society of the
  Pacific Conference Series Vol. 527, Astronomical Data Analysis Software and
  Systems XXIX. p.~635 (\mn@eprint {arXiv} {2006.02955})

\bibitem[\protect\citeauthoryear{{Kuiack}, {Wijers}, {Shulevski}, {Rowlinson},
  {Huizinga}, {Molenaar}  \& {Prasad}}{{Kuiack} et~al.}{2021}]{kuiack2021}
{Kuiack} M.,  {Wijers} R. A.~M.~J.,  {Shulevski} A.,  {Rowlinson} A.,
  {Huizinga} F.,  {Molenaar} G.,   {Prasad} P.,  2021, \mn@doi [\mnras]
  {10.1093/mnras/stab1504}, \href
  {https://ui.adsabs.harvard.edu/abs/2021MNRAS.505.2966K} {505, 2966}

\bibitem[\protect\citeauthoryear{{Levinson}, {Ofek}, {Waxman}  \&
  {Gal-Yam}}{{Levinson} et~al.}{2002}]{levinson2002}
{Levinson} A.,  {Ofek} E.~O.,  {Waxman} E.,   {Gal-Yam} A.,  2002, \mn@doi
  [\apj] {10.1086/341866}, \href
  {https://ui.adsabs.harvard.edu/abs/2002ApJ...576..923L} {576, 923}

\bibitem[\protect\citeauthoryear{{Lorimer}, {Bailes}, {McLaughlin}, {Narkevic}
  \& {Crawford}}{{Lorimer} et~al.}{2007}]{lorimer2007}
{Lorimer} D.~R.,  {Bailes} M.,  {McLaughlin} M.~A.,  {Narkevic} D.~J.,
  {Crawford} F.,  2007, \mn@doi [Science] {10.1126/science.1147532}, \href
  {https://ui.adsabs.harvard.edu/abs/2007Sci...318..777L} {318, 777}

\bibitem[\protect\citeauthoryear{{Mauch} et~al.,}{{Mauch}
  et~al.}{2020}]{2020ApJ...888...61M..Mauch2020}
{Mauch} T.,  et~al., 2020, \mn@doi [\apj] {10.3847/1538-4357/ab5d2d}, \href
  {https://ui.adsabs.harvard.edu/abs/2020ApJ...888...61M} {888, 61}

\bibitem[\protect\citeauthoryear{{Mooley}, {Frail}, {Ofek}, {Miller},
  {Kulkarni}  \& {Horesh}}{{Mooley} et~al.}{2013}]{mooley2013}
{Mooley} K.~P.,  {Frail} D.~A.,  {Ofek} E.~O.,  {Miller} N.~A.,  {Kulkarni}
  S.~R.,   {Horesh} A.,  2013, \mn@doi [\apj] {10.1088/0004-637X/768/2/165},
  \href {https://ui.adsabs.harvard.edu/abs/2013ApJ...768..165M} {768, 165}

\bibitem[\protect\citeauthoryear{{Mooley} et~al.,}{{Mooley}
  et~al.}{2016}]{mooley2016}
{Mooley} K.~P.,  et~al., 2016, \mn@doi [\apj] {10.3847/0004-637X/818/2/105},
  \href {https://ui.adsabs.harvard.edu/abs/2016ApJ...818..105M} {818, 105}

\bibitem[\protect\citeauthoryear{{Murphy} et~al.,}{{Murphy}
  et~al.}{2021}]{murphy2021}
{Murphy} T.,  et~al., 2021, \mn@doi [\pasa] {10.1017/pasa.2021.44}, \href
  {https://ui.adsabs.harvard.edu/abs/2021PASA...38...54M} {38, e054}

\bibitem[\protect\citeauthoryear{{Offringa}, {de Bruyn}, {Biehl}, {Zaroubi},
  {Bernardi}  \& {Pandey}}{{Offringa} et~al.}{2010}]{offringa2010}
{Offringa} A.~R.,  {de Bruyn} A.~G.,  {Biehl} M.,  {Zaroubi} S.,  {Bernardi}
  G.,   {Pandey} V.~N.,  2010, \mn@doi [\mnras]
  {10.1111/j.1365-2966.2010.16471.x}, \href
  {https://ui.adsabs.harvard.edu/abs/2010MNRAS.405..155O} {405, 155}

\bibitem[\protect\citeauthoryear{{Offringa} et~al.,}{{Offringa}
  et~al.}{2014}]{2014MNRAS.444..606O..Offringa2014..WSClean}
{Offringa} A.~R.,  et~al., 2014, \mn@doi [\mnras] {10.1093/mnras/stu1368},
  \href {https://ui.adsabs.harvard.edu/abs/2014MNRAS.444..606O} {444, 606}

\bibitem[\protect\citeauthoryear{{Petroff}, {Hessels}  \& {Lorimer}}{{Petroff}
  et~al.}{2022}]{petroff2022}
{Petroff} E.,  {Hessels} J.~W.~T.,   {Lorimer} D.~R.,  2022, \mn@doi [\aapr]
  {10.1007/s00159-022-00139-w}, \href
  {https://ui.adsabs.harvard.edu/abs/2022A&ARv..30....2P} {30, 2}

\bibitem[\protect\citeauthoryear{{Pietka}, {Fender}  \& {Keane}}{{Pietka}
  et~al.}{2015}]{2015MNRAS.446.3687P..Pietka2015}
{Pietka} M.,  {Fender} R.~P.,   {Keane} E.~F.,  2015, \mn@doi [\mnras]
  {10.1093/mnras/stu2335}, \href
  {https://ui.adsabs.harvard.edu/abs/2015MNRAS.446.3687P} {446, 3687}

\bibitem[\protect\citeauthoryear{{Prasad} et~al.,}{{Prasad}
  et~al.}{2016}]{2016JAI.....541008P..Prasad2016..AARTFAAC}
{Prasad} P.,  et~al., 2016, \mn@doi [Journal of Astronomical Instrumentation]
  {10.1142/S2251171716410087}, \href
  {https://ui.adsabs.harvard.edu/abs/2016JAI.....541008P} {5, 1641008}

\bibitem[\protect\citeauthoryear{{Rowlinson} et~al.,}{{Rowlinson}
  et~al.}{2016}]{2016MNRAS.458.3506R..Rowlinson2016}
{Rowlinson} A.,  et~al., 2016, \mn@doi [\mnras] {10.1093/mnras/stw451}, \href
  {https://ui.adsabs.harvard.edu/abs/2016MNRAS.458.3506R} {458, 3506}

\bibitem[\protect\citeauthoryear{{Rowlinson} et~al.,}{{Rowlinson}
  et~al.}{2022}]{rowlinson2022}
{Rowlinson} A.,  et~al., 2022, \mn@doi [\mnras] {10.1093/mnras/stac2460}, \href
  {https://ui.adsabs.harvard.edu/abs/2022MNRAS.517.2894R} {517, 2894}

\bibitem[\protect\citeauthoryear{{Serra} et~al.,}{{Serra}
  et~al.}{2019}]{2019A&A...628A.122S..Serra2019}
{Serra} P.,  et~al., 2019, \mn@doi [\aap] {10.1051/0004-6361/201936114}, \href
  {https://ui.adsabs.harvard.edu/abs/2019A&A...628A.122S} {628, A122}

\bibitem[\protect\citeauthoryear{{Skrutskie} et~al.,}{{Skrutskie}
  et~al.}{2006}]{2006AJ....131.1163S..Skrutskie2006..2MASS}
{Skrutskie} M.~F.,  et~al., 2006, \mn@doi [\aj] {10.1086/498708}, \href
  {https://ui.adsabs.harvard.edu/abs/2006AJ....131.1163S} {131, 1163}

\bibitem[\protect\citeauthoryear{{Stewart} et~al.,}{{Stewart}
  et~al.}{2016}]{stewart2016}
{Stewart} A.~J.,  et~al., 2016, \mn@doi [\mnras] {10.1093/mnras/stv2797}, \href
  {https://ui.adsabs.harvard.edu/abs/2016MNRAS.456.2321S} {456, 2321}

\bibitem[\protect\citeauthoryear{{Swinbank} et~al.,}{{Swinbank}
  et~al.}{2015}]{2015A&C....11...25S..Swinbank2015}
{Swinbank} J.~D.,  et~al., 2015, \mn@doi [Astronomy and Computing]
  {10.1016/j.ascom.2015.03.002}, \href
  {https://ui.adsabs.harvard.edu/abs/2015A&C....11...25S} {11, 25}

\bibitem[\protect\citeauthoryear{{Thyagarajan}, {Helfand}, {White}  \&
  {Becker}}{{Thyagarajan} et~al.}{2011}]{thyagarajan2011}
{Thyagarajan} N.,  {Helfand} D.~J.,  {White} R.~L.,   {Becker} R.~H.,  2011,
  \mn@doi [\apj] {10.1088/0004-637X/742/1/49}, \href
  {https://ui.adsabs.harvard.edu/abs/2011ApJ...742...49T} {742, 49}

\bibitem[\protect\citeauthoryear{{Virtanen} et~al.,}{{Virtanen}
  et~al.}{2020}]{2020NatMe..17..261V..Virtanen2020.scipy}
{Virtanen} P.,  et~al., 2020, \mn@doi [Nature Methods]
  {10.1038/s41592-019-0686-2}, \href
  {https://ui.adsabs.harvard.edu/abs/2020NatMe..17..261V} {17, 261}

\bibitem[\protect\citeauthoryear{{Wang} et~al.,}{{Wang}
  et~al.}{2022}]{2022MNRAS.516.5972W..Wang2022}
{Wang} Z.,  et~al., 2022, \mn@doi [\mnras] {10.1093/mnras/stac2542}, \href
  {https://ui.adsabs.harvard.edu/abs/2022MNRAS.516.5972W} {516, 5972}

\bibitem[\protect\citeauthoryear{{Wang} et~al.,}{{Wang}
  et~al.}{2023}]{2023MNRAS.tmp.1665W..Wang2023}
{Wang} Y.,  et~al., 2023, \mn@doi [\mnras] {10.1093/mnras/stad1727}, \href
  {https://ui.adsabs.harvard.edu/abs/2023MNRAS.tmp.1665W} {}

\bibitem[\protect\citeauthoryear{{Yu}, {Zijlstra}  \& {Jiang}}{{Yu}
  et~al.}{2021}]{Yu2021}
{Yu} B.,  {Zijlstra} A.,   {Jiang} B.,  2021, \mn@doi [Universe]
  {10.3390/universe7050119}, \href
  {https://ui.adsabs.harvard.edu/abs/2021Univ....7..119Y} {7, 119}

\bibitem[\protect\citeauthoryear{{de Blok} et~al.,}{{de Blok}
  et~al.}{2016}]{2016mks..confE...7D..deBlok2016}
{de Blok} W.~J.~G.,  et~al., 2016, in MeerKAT Science: On the Pathway to the
  SKA. p.~7 (\mn@eprint {arXiv} {1709.08458})

\bibitem[\protect\citeauthoryear{{de Blok} et~al.,}{{de Blok}
  et~al.}{2020}]{2020A&A...643A.147D..deBlok2020}
{de Blok} W.~J.~G.,  et~al., 2020, \mn@doi [\aap]
  {10.1051/0004-6361/202038894}, \href
  {https://ui.adsabs.harvard.edu/abs/2020A&A...643A.147D} {643, A147}

\bibitem[\protect\citeauthoryear{{de Ruiter}, {Leseigneur}, {Rowlinson},
  {Wijers}, {Drabent}, {Intema}  \& {Shimwell}}{{de Ruiter}
  et~al.}{2021}]{2021MNRAS.508.2412D..deRuiter2021}
{de Ruiter} I.,  {Leseigneur} G.,  {Rowlinson} A.,  {Wijers} R. A.~M.~J.,
  {Drabent} A.,  {Intema} H.~T.,   {Shimwell} T.~W.,  2021, \mn@doi [\mnras]
  {10.1093/mnras/stab2695}, \href
  {https://ui.adsabs.harvard.edu/abs/2021MNRAS.508.2412D} {508, 2412}

\makeatother
\end{thebibliography}




\appendix
\section{Filtering transient candidates}
\label{sec:extrafigures}

\begin{figure}
    \centering
    \includegraphics[width=\columnwidth]{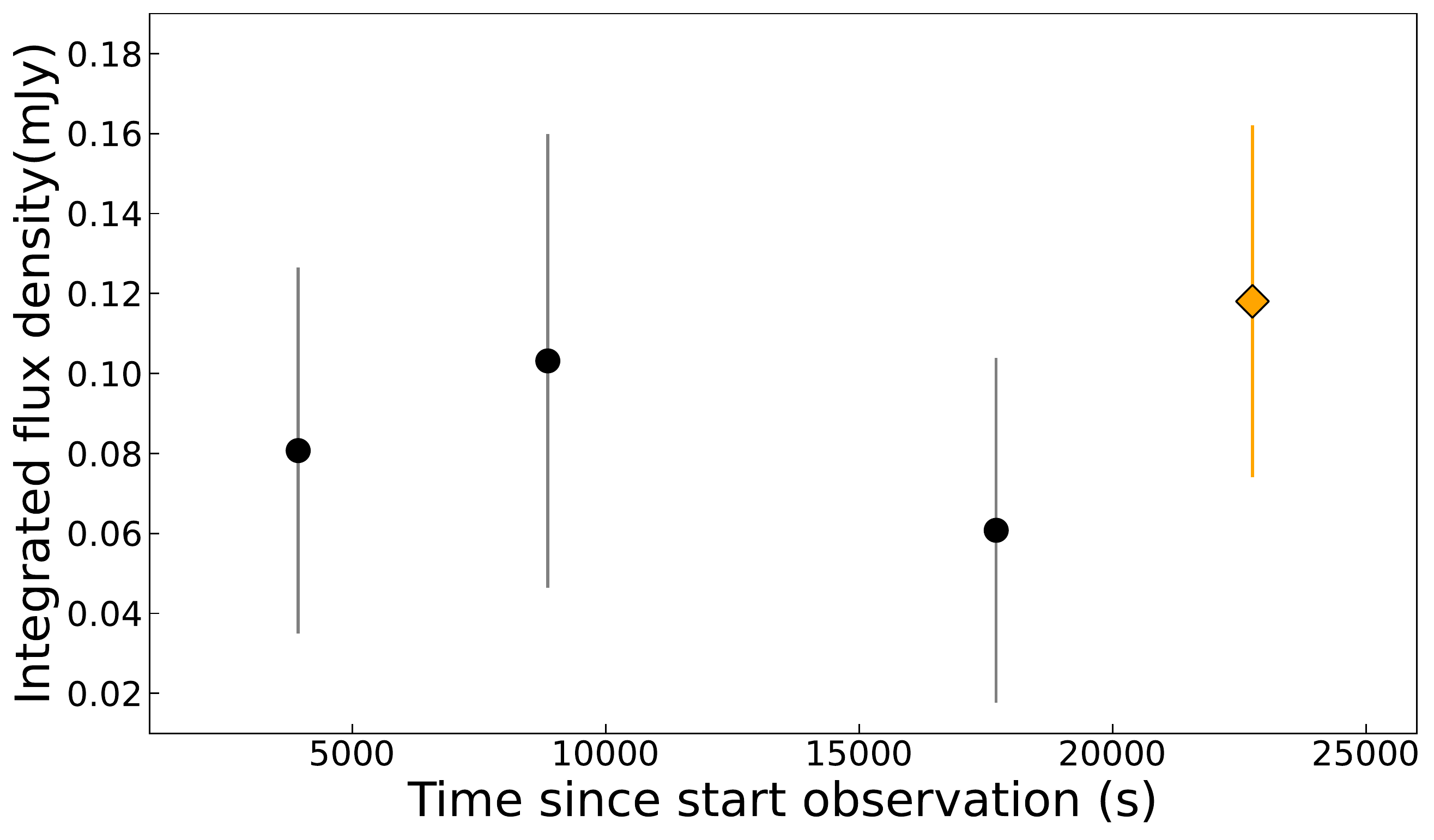}
    \caption{An example light curve of a rejected candidate detected in the 1 hour time scale data, where the {\sc TraP} detection is shown with the yellow diamond. The candidate was detected with a signal-to-noise of 5.3$\sigma$, but from the light curve we deduce that the variation is not significant, and is likely caused by an artefact in the image. Therefore, this candidate is rejected.}
    \label{fig:lc_detec_threshold}
\end{figure}

\begin{figure}
    \centering
    \includegraphics[width=\columnwidth]{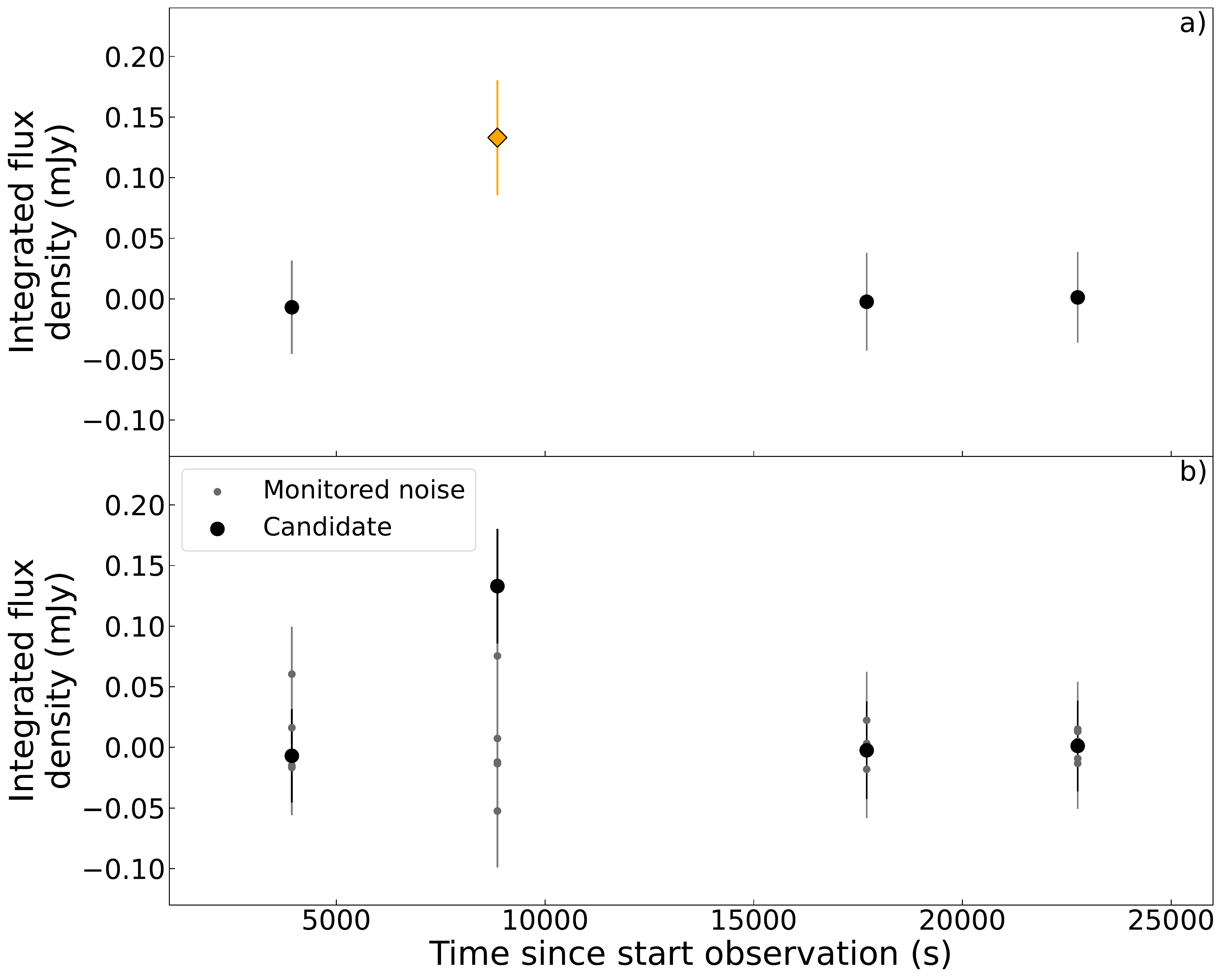}
    \caption{An example light curve of a rejected candidate detected in the 1 hour time scale. Panel (a) shows the light curve from the candidate where the {\sc TraP} detection is shown with the yellow diamond. Here the variation of the detection appears significant. Panel (b) shows the light curve of the candidate, as well as light curves of 5 noise points which were randomly selected in the vicinity (see Section \ref{sec:filtering}). From these light curves, we deduce that the variation of the candidate is not significant when compared with the trends in the surrounding noise. Therefore, this candidate is rejected.}
    \label{fig:lc_noise_peak}
\end{figure}

\begin{figure}
    \centering
    \includegraphics[width=0.98\columnwidth]{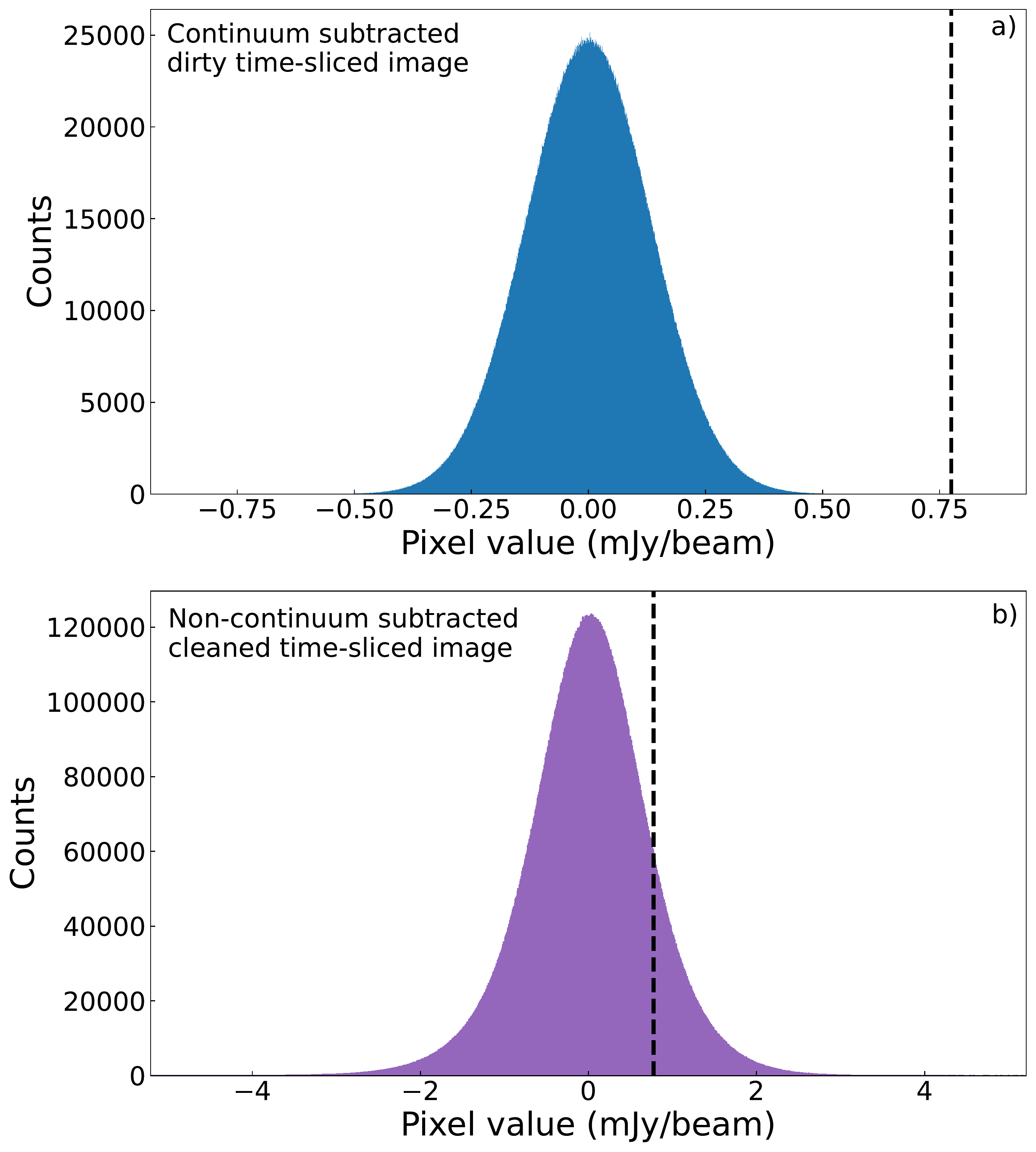}
    \caption{The pixel value distributions of an example image (image 39) from the 128 second data set. Panel (a) shows the distribution for the dirty continuum subtracted data, and panel (b) for the cleaned non-continuum subtracted data. The black dotted line shows the pixel value for the 5.8$\sigma$ detection threshold as determined for the dirty continuum subtracted image. Therefore, we find that any candidate detected in the dirty continuum subtracted time-sliced image at a signal-to-noise of 5.8$\sigma$, would likely not be significantly detected in the cleaned non-continuum time-sliced image.}
    \label{fig:noise_cleaned_images}
\end{figure}

\section{Previous transient surveys}
\label{sec:surveys}

\begin{table*}
\centering
\begin{tabular}{|c c c l|}
\hline
Sensitivity & Transient Surface & Time-scale & Author \\
(Jy) & Density (deg$^{-2}$) & (days) & \\
\hline
$1.5 \times 10^{-4}$ & $<5.7$                 & $19$                 & \cite{carilli2003} \\
$7 \times 10^{-2}$   & $<3 \times 10^{-3}$    & $1$                  & \cite{bower2011} \\
$3$                  & $<9 \times 10^{-4}$    & $1$                  & \cite{bower2011} \\
$0.35$               & $<6 \times 10^{-4}$    & $1$                  & \cite{croft2011} \\
$8 \times 10^{-3}$   & $<3.2 \times 10^{-2}$  & $4$                  & \cite{bell2011} \\
$1 \times 10^{-3}$   & $6.5 \times 10^{-3}$   & $2 \times 10^{-3}$   & \cite{thyagarajan2011} \\
$3.7 \times 10^{-4}$ & $<0.6$                 & $1.4 \times 10^{-2}$ & \cite{frail2012} \\
$2.1 \times 10^{-4}$ & $<0.37$                & $1$                  & \cite{mooley2013} \\
$3$                  & $2 \times 10^{-6}$     & $1$                  & \cite{aoki2014} \\
$5 \times 10^{-4}$   & $<17$                  & $2 \times 10^{-2}$   & \cite{alexander2015} \\
$6 \times 10^{-4}$   & $<8 \times 10^{-2}$    & $1$                  & \cite{mooley2016} \\
$1.5 \times 10^{-3}$ & $<0.3$                 & $1$                  & \cite{bhandari2018} \\
$1.5 \times 10^{-3}$ & $<0.3$                 & $12$                 & \cite{bhandari2018} \\
$1 \times 10^{-3}$   & $<3.7 \times 10^{-2}$  & $7$                  & \cite{rowlinson2022} \\
$5.64 \times 10^{-2}$ & $<6.7 \times 10^{-5}$ & $9.3\times10^{-5}$   & This Work (8 seconds)\\
$1.92 \times 10^{-2}$ & $<1.1 \times 10^{-3}$ & $1.5\times10^{-3}$   & This Work (128 seconds)\\
$3.9 \times 10^{-3}$  & $<3.2 \times 10^{-2}$ & $4.2\times10^{-2}$   & This Work (1 hour)\\

\hline
\end{tabular}
\caption[\url{http://www.tauceti.caltech.edu/kunal/radio-transient-surveys/index.html}]{The transient surface density constraints at ${\approx}1.4$ GHz obtained using a range of transient surveys for transients with time-scales of less than 1 month. These data are used to plot Figure \ref{fig:surfaceDensityPlot} and were obtained from \citet{mooley2016}. \protect\footnotemark}
\label{table:transientSDs}
\end{table*}

\newpage
\footnotetext{\url{http://www.tauceti.caltech.edu/kunal/radio-transient-surveys/index.html}}


\bsp	
\label{lastpage}
\end{document}